\documentclass[10pt, prd,twocolumn, nofootinbib,preprint,superscriptaddress]{revtex4}

\usepackage[T1]{fontenc}
\usepackage{amsmath,amssymb}
\usepackage{epsfig}
\usepackage{float}
\usepackage{graphicx}
\usepackage[colorlinks,citecolor=blue]{hyperref}
\usepackage{pdfpages}
\usepackage{color}

\begin{document}
	\title{Singlet-Doublet Fermion Origin of Dark Matter, Neutrino Mass and W-Mass Anomaly}
	\author{Debasish Borah}
	\email{dborah@iitg.ac.in}
	\affiliation{Department of Physics, Indian Institute of Technology Guwahati, Assam 781039, India}
	
	\author{Satyabrata Mahapatra}
	\email{ph18resch11001@iith.ac.in}
	\affiliation{Department of Physics, Indian Institute of Technology Hyderabad, Kandi, Sangareddy 502285, Telangana, India}
	
	\author{Narendra Sahu}
	\email{nsahu@phy.iith.ac.in}
	\affiliation{Department of Physics, Indian Institute of Technology Hyderabad, Kandi, Sangareddy 502285, Telangana, India}
	
	\begin{abstract}
	Motivated by the recently reported anomaly in W boson mass by the CDF collaboration with $7\sigma$ statistical significance, we consider a singlet-doublet (SD) Majorana fermion dark matter (DM) model where the required correction to W boson mass arises from radiative corrections induced by SD fermions. While a single generation of SD fermions, odd under an unbroken $Z_2$ symmetry, can not explain the W boson mass anomaly while being consistent with DM phenomenology, two generations of SD fermions can do so with the heavier generation playing the dominant role in W-mass correction and lighter generation playing the role in DM phenomenology. Additionally, such multiple generations of SD fermions can also generate light neutrino masses radiatively if a $Z_2$-odd singlet scalar is included.
	\end{abstract}
	
	\maketitle
	\noindent

\noindent
\textbf{\emph{Introduction}:} The recent announcement of the updated measurement of the W boson mass $M_W = 80433.5 \pm 9.4$ MeV \cite{CDF:2022hxs} by CDF collaboration at Fermilab using the data corresponding to 8.8 ${\rm fb}^{-1}$ integrated luminosity collected at the CDF-II detector of Fermilab Tevatron collider has led to a $7\sigma$ discrepancy with the standard model (SM) expectation ($M_{W}=80357\pm6$ MeV). Possible implications and interpretations of this anomaly have been discussed in several recent works. For example, connections to effective field theory \cite{Fan:2022yly, Bagnaschi:2022whn}, electroweak precision parameters \cite{deBlas:2022hdk, Strumia:2022qkt, Asadi:2022xiy, Lu:2022bgw, Carpenter:2022oyg}, beyond standard model (BSM) physics like dark matter (DM) \cite{Fan:2022dck, Zhu:2022tpr, Zhu:2022scj, Kawamura:2022uft, Nagao:2022oin, Zhang:2022nnh, Liu:2022jdq}, additional scalar fields \cite{Sakurai:2022hwh, Cacciapaglia:2022xih, Song:2022xts, Bahl:2022xzi, Cheng:2022jyi, Babu:2022pdn, Heo:2022dey, Ahn:2022xeq, Zheng:2022irz, Perez:2022uil, Kanemura:2022ahw, Borah:2022obi, Popov:2022ldh, Arcadi:2022dmt, Ghorbani:2022vtv,Han:2022juu}, supersymmetry \cite{Du:2022pbp, Tang:2022pxh,Yang:2022gvz, Athron:2022isz,Ghoshal:2022vzo} and several others \cite{Athron:2022qpo, Blennow:2022yfm,Heckman:2022the, Lee:2022nqz, DiLuzio:2022xns,Paul:2022dds,Biekotter:2022abc,Balkin:2022glu,Cheung:2022zsb,Du:2022brr, Endo:2022kiw, Crivellin:2022fdf, Mondal:2022xdy, Chowdhury:2022moc, Du:2022fqv, Bhaskar:2022vgk, Yuan:2022cpw, Arias-Aragon:2022ats} have already been outlined. Motivated by this and assuming that the reported CDF-II anomaly is purely due to BSM physics, we consider a singlet-doublet (SD) Majorana fermion dark matter scenario to explain this anomaly. We find that a single generation of SD fermion, stabilised by an unbroken $Z_2$ symmetry, can not explain W-mass anomaly while being consistent with DM phenomenology, irrespective of Dirac or Majorana nature of DM. Due to possibility of connecting to origin of light neutrino masses, we consider the Majorana nature of SD DM here and show that additional $Z_2$-odd SD fermions can lead to successful explanation of W-mass anomaly with the additional advantage of explaining non-zero neutrino mass and mixing if an extra $Z_2$-odd singlet scalar is incorporated leading to a one-loop neutrino mass diagram. In such a setup, the heavier SD fermions play dominant role in W-mass correction while the lightest one remains consistent with DM phenomenology.

\vspace{0.2cm}
\noindent 
\textbf{\emph{The Model}:} 
Singlet-doublet fermion DM has been studied extensively in the literature \cite{Mahbubani:2005pt,DEramo:2007anh,Enberg:2007rp,Cohen:2011ec,Cheung:2013dua, Restrepo:2015ura,Calibbi:2015nha,Cynolter:2015sua, Bhattacharya:2015qpa,Bhattacharya:2017sml, Bhattacharya:2018fus,Bhattacharya:2018cgx,DuttaBanik:2018emv,Barman:2019tuo,Bhattacharya:2016rqj, Calibbi:2018fqf, Barman:2019aku, Dutta:2020xwn, Borah:2021khc, Borah:2021rbx} in different contexts. We extend the SM by a vector like fermion doublet $\Psi$ and a Majorana singlet fermion $N_R$, both odd under an unbroken $Z_2$ symmetry. The Lagrangian of the model is given by
\begin{eqnarray}
\label{model_Lagrangian}
 \mathcal{L} &=& \mathcal{L}_{\rm SM} + \overline{\Psi_i} \left( i\gamma^\mu D_\mu - M_i \right) \Psi_i +\overline{{N}_{R_i}} i\gamma^\mu\partial_\mu {N}_{R_i}\nonumber \\&-& \frac{1}{2}M_{R_i} \overline{{N}_{R_i}^c} {N}_{R_i} + \mathcal{L}_{\rm yuk}.    \end{eqnarray}
Here $\mathcal{L}_{\rm SM}$ denotes the SM Lagrangian and the Yukawa Lagrangian $\mathcal{L}_{\rm yuk}$ of the newly introduced fields plays the key role in DM phenomenology. It is given by
\begin{equation}
    -\mathcal{L}_{\rm yuk} = \frac{Y_i}{\sqrt{2}}\overline{\Psi_{i}}\Tilde{H}\big(N_{R_i}+(N_{R_i})^c\big) + Y_{j \alpha }\overline{\psi_{R_j}} \phi L_{\alpha} + {\rm h.c.}
    \label{eq:yukawa}
\end{equation}
where $H$ is the SM Higgs doublet and $L_{\alpha}$ denotes usual lepton doublets. The subscript $i$ denotes the number of generations which, to begin with, is assumed to be just one for the sake of minimality. The mass terms of these fields for a single generation, after electroweak symmetry breaking, can then be written together as
\begin{eqnarray}
  -\mathcal{L}_{\rm mass} &=& M_i\overline{\psi^0_{L_i}}\psi^0_{R_i} + \frac{1}{2}M_{R_i}\overline{N}_{R_i}(N_{R_i})^c \nonumber \\&+& \frac{M_{D_i}}{\sqrt{2}} (\overline{\psi^0_{L_i}}N_{R_i}+\overline{\psi^0_{R_i}}(N_{R_i})^c) + {\rm h.c.} 
  \label{l_mass}
\end{eqnarray}
where $M_{D_i}={Y_i v }/{\sqrt{2}}$, where $ v = 246$ GeV is the vacuum expectation value (VEV) of the SM Higgs doublet's neutral component. Writing these mass terms in the basis 
$ ((\psi^0_R)^c, \psi^0_L, (N_{R_1})^c)^T$, we get the following mass matrix
\begin{eqnarray}\label{dark-sector-mass}
\mathcal{M}_i=
\left(
\begin{array}{ccc}
0 &M_i &{M_{D_i}}/{\sqrt{2}}\\
M_i &0 &{M_{D_i}}/{\sqrt{2}}\\
{M_{D_i}}/{\sqrt{2}} &{M_{D_i}}/{\sqrt{2}} &M_{R_i}\\
\end{array}
\right)\,.
\end{eqnarray}
The mass matrix of this texture can be diagonalised by a single unitary matrix 
$\mathcal{U (\theta)}=U_{13}(\theta_{13}=\theta).U_{23}(\theta_{23}=0).U_{12}(\theta_{12}=\frac{\pi}{4})$, which is essentially characterised by a single angle $\theta_{13}=\theta$. So we diagonalise the mass matrix $\mathcal{M}_i$ for each singlet-doublet generation by $\mathcal{U}.\mathcal{M}_i.\mathcal{U}^T = \mathcal{M}^{\rm diag}_i$, 
where the unitary matrix $\mathcal{U}$ has the form
\begin{equation}
\label{diagonalizing_matrix}
\mathcal{U}= \left(
\begin{array}{ccc}
1 & 0 & 0\\
0 & e^{i\pi/2} & 0\\
0 & 0 & 1\\
\end{array}
\right)
\left(
\begin{array}{ccc}
\frac{1}{\sqrt{2}}\cos\theta_i &\frac{1}{\sqrt{2}}\cos\theta_i &\sin\theta_i\\
-\frac{1}{\sqrt{2}} &\frac{1}{\sqrt{2}} &0\\
-\frac{1}{\sqrt{2}}\sin\theta_i &-\frac{1}{\sqrt{2}}\sin\theta_i &\cos\theta_i\\
\end{array}
\right)\\,
\end{equation}
The extra phase matrix is multiplied to make sure all the mass eigenvalues are positive and the mixing angle $\theta_i$ for every generation is given by
\begin{equation}
\tan2\theta_i = \frac{2 M_{D_i}}{M_i-M_{R_i}}.
\label{sdmixing}
\end{equation}
 
The physical states that emerge from the above diagonalisation are defined as $\chi_{ik}=\frac{\chi_{_{ikL}}+(\chi_{_{ikL}})^c}{\sqrt{2}}~(k=1,2,3)$ and are 
related to the unphysical or flavour states as
\begin{equation}
\begin{aligned}
\chi_{_{i1L}} & = \frac{\cos\theta_i}{\sqrt{2}}( \psi^0_{L_i}+(\psi^0_{R_i})^c  )+\sin\theta_i (N_{R_i})^c,
\\
\chi_{_{i2L}} & =  \frac{i}{\sqrt{2}}(\psi^0_{L_i} - (\psi^0_{R_i})^c), 
\\
\chi_{_{i3L}} & =  -\frac{\sin\theta_i}{\sqrt{2}}(\psi^0_{L_i} + (\psi^0_{R_i})^c ) +\cos\theta_i(N_{R_i})^c \,.
\end{aligned}
\end{equation} 
All the three physical states for each generation $i$ $\chi_{i_1}, \chi_{i_2} ~{\rm and} ~\chi_{i_3}$ are therefore of Majorana nature and their 
mass eigenvalues can be expressed respectively as,
\begin{equation}\label{darkmass}
\begin{aligned}
M_{\chi_{i1}} & = M_i \cos^2\theta_i + M_{R_i} \sin^2\theta_i + M_{D_i}\sin2\theta_i,
\\
M_{\chi_{i2}} & = M_i,
\\
M_{\chi_{i3}} & = M_{R_i} \cos^2\theta_i + M_i\sin^2\theta_i - M_{D_i}\sin2\theta_i\,.
\\
\end{aligned}
\end{equation}
Using the relation $\mathcal{U}.\mathcal{M}_i.\mathcal{U}^T = \mathcal{M}^{\rm diag.}_i$, one can express $Y_{1i}$, $M_i$ and $M_{R_i}$ in 
terms of the physical masses and the mixing angle as,
\begin{equation}
\begin{aligned}
Y_{i} & = \frac{\sqrt{2}~\Delta M_i ~\sin2\theta_i}{v},
\\
M_i & = M_{\chi_{i1}} \cos^2\theta_i +M_{\chi_{i3}} \sin^2\theta_i, 
\\
M_{R_i} & = M_{\chi_{i3}} \cos^2\theta_i +  M_{\chi_{i1}}\sin^2\theta_i; 
\\
\end{aligned}
\label{y1}
\end{equation}
where $\Delta M_i=(M_{\chi_{i1}}-M_{\chi_{i3}})$. The phenomenology of dark sector is therefore governed mainly by the three independent parameters, DM mass, mass splitting between DM and the heavier neutral component, and doublet-singlet mixing given in Eq. \eqref{sdmixing}. Here it is worth mentioning that the mass of $\chi_{i 2}$ is exactly same as the mass of the charged fermion $\psi^{-}_i$ {\it i.e.} $M_{\chi_{i 2}}=M_{\psi^{-}_i} = M_{i}$.

\begin{figure}[h!]
\includegraphics[scale=0.5]{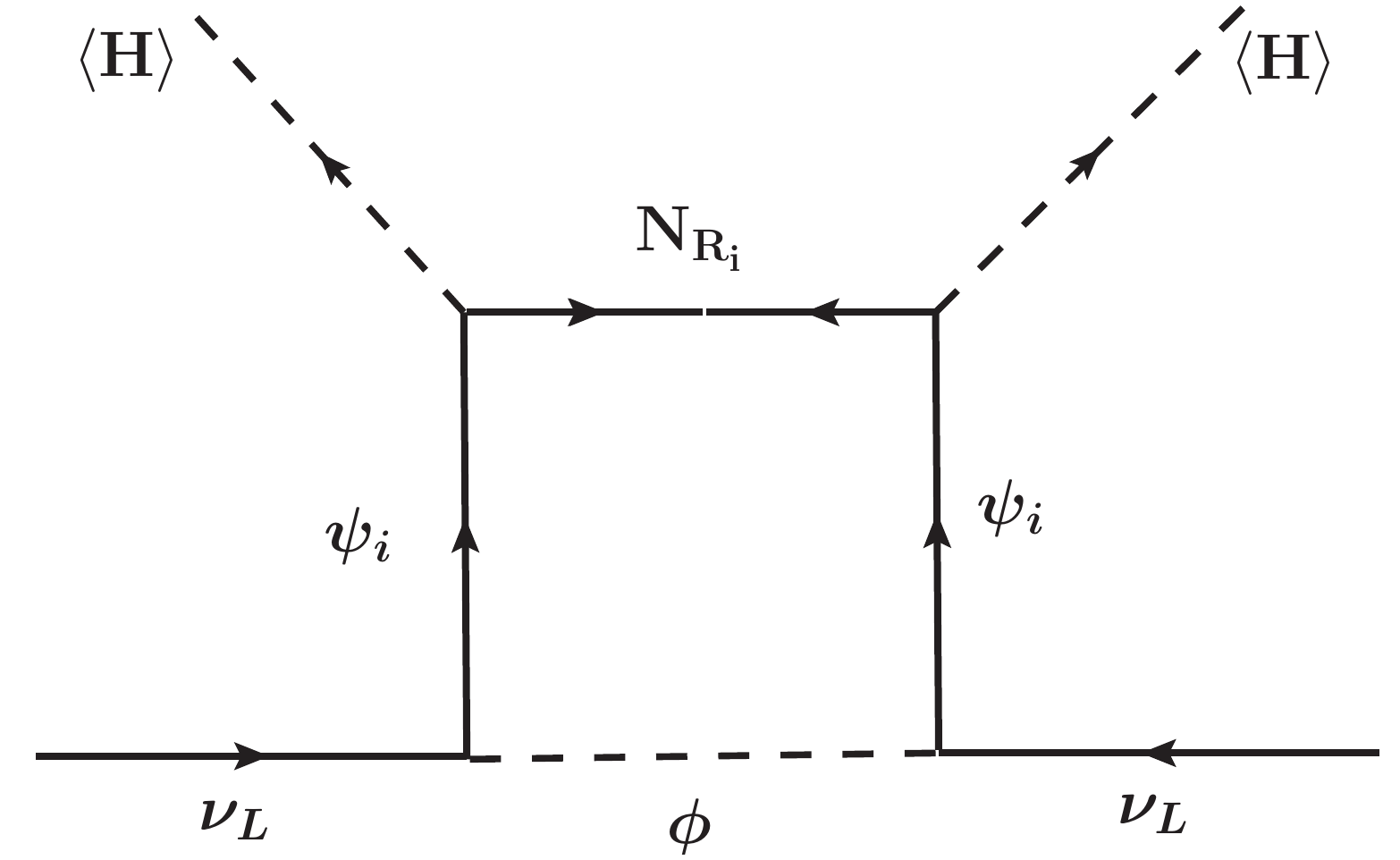}
\caption{Radiative Generation of neutrino mass with dark sector particles in the loop.}\label{numass}
\end{figure}

For at least two generations of singlet-doublet fermions and an additional $Z_2$-odd singlet scalar $\phi$, the light neutrino mass in our setup can be generated at one-loop level with the dark sector particles in the loop as shown in the Fig.~\ref{numass}. The corresponding one-loop expression for light neutrino masses can be obtained following earlier related works \cite{Fraser:2014yha,Konar:2020wvl}. 

\vspace{0.2cm}
\noindent
\textbf{\emph{Dark Matter Phenomenology}:}  With the details of particle spectrum and couplings discussed above, we then numerically calculate the DM abundance using the package \texttt{micrOmega}~\cite{Belanger:2008sj}.
\begin{figure*}[!htb]
\includegraphics[scale=0.5]{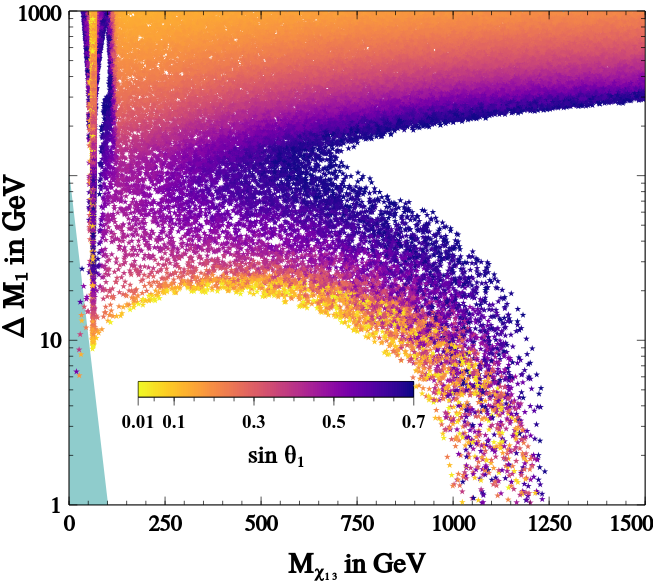}
\includegraphics[scale=0.5]{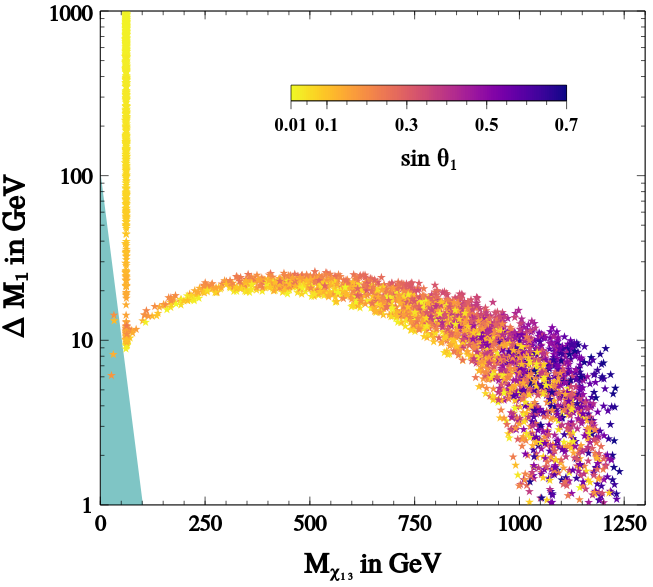}
\caption{[Left panel]:Parameter space satisfying the correct relic density in the plane of $M_{\chi_{1 3}}$ and $\Delta M$ with $\sin\theta$ shown by the colour map.
[Right panel]: Parameter space consistent with both relic density (from Planck) and Direct Search constraints (from XENON1T) in the $\Delta M $ versus $M_{\chi_3}$. The dark cyan shaded region in the bottom left corner is ruled out by the LEP exclusion bound on charged fermion mass $M_{\psi^\pm}=M_2>102.7$GeV.}
\label{DMfig}
\end{figure*}

In the left panel of Fig.~\ref{DMfig}, the parameter space giving rise to correct relic density \cite{Aghanim:2018eyx} has been shown in the plane of $\Delta M_1$ versus $M_{\chi_{1 3}}$ with the singlet-doublet mixing $\sin\theta_1$, indicated by the colour map. The bifurcation of the allowed parameter space around $\Delta M_1 \sim 50$ GeV can be clearly explained by separating the plane into two regions as : (I) the bottom portion with small $\Delta M$ ($\Delta M \leq$ 50 GeV), where with increase in DM mass ($M_{\chi_{1 3}}$), $\Delta M$ decreases and (II) the top portion with large $\Delta M$ ($\Delta M \geq$ 50 GeV), where $\Delta M$ increases slowly with larger DM mass $M_{\chi_{1 3}}$. In the first region, for a fixed $\sin\theta_1$, with increase in DM mass $M_{\chi_{1 3}}$, the annihilation cross-section decreases, and hence it requires more co-annihilation contributions to get the correct relic density, resulting in $\Delta M$ to decrease. Hence the region below this coloured zone corresponds to under-abundant DM. Similarly the region above this coloured region corresponds to over-abundant DM due to the same logic. In this region the Yukawa coupling $Y_1$ ( $Y_1 \propto \Delta M_1 \sin 2\theta_1$) is comparatively small as $\Delta M_1$ is small and hence the annihilation cross-section is small. In addition as the annihilation cross-section decreases with increase in DM mass, when DM mass is sufficiently heavy ($M_{\chi_{1 3}} > 1.2$ TeV), annihilation contribution to relic becomes extremely small to be compensated by the co-annihilation even when $\Delta M \rightarrow 0$, and thus results in DM over-abundance.
	
 In the second region (II), as $\Delta M$ is large, the co-annihilation contribution to relic is negligible and hence the Higgs mediated annihilation processes dominantly decide the relic density. As Higgs Yukawa coupling $Y_1 \propto \Delta M \sin2\theta_1$, for a given $\sin\theta_1$, if $\Delta M$ increases then it increases $Y_1$ and hence larger annihilation cross-section which results in DM under-abundance. Thus it can only be brought back to the correct ballpark by having a larger DM mass. Similarly, larger $\sin\theta_1$ requires smaller $\Delta M_1$. Therefore, the region above the coloured zone is under-abundant, while the region below this coloured zone is over-abundant.

	Now, if relic density allowed parameter space is confronted with the constraints from DM direct search experiments XENON1T~\cite{Aprile:2018dbl}, then we see that the parameter space is crucially tamed down as shown in the right panel of Fig.~\ref{DMfig}. Here the elastic scattering of the DM off nuclei occur via SM Higgs mediation. It is worth mentioning that, the absence of tree level Z-mediated DM-nucleon scattering makes a crucial difference in the direct search allowed parameter space as compared to singlet-doublet Dirac fermion DM elaborated in~\cite{Bhattacharya:2017sml,Barman:2019tuo,Bhattacharya:2018cgx,Bhattacharya:2016rqj,Bhattacharya:2015qpa,Bhattacharya:2018fus,Borah:2022obi}. In the present scenario, a large $\sin \theta_1$ is allowed from both relic as well as direct search constraints, whereas in case of singlet-doublet Dirac fermion DM, only upto $\sin \theta_1 \sim 0.01$ is allowed. The spin-independent (SI) DM-nucleon interaction cross-section mediated by the SM Higgs can be written as~\cite{Dutta:2020xwn}
	\begin{equation}
		\label{dda2}
		\begin{aligned}
			\sigma_{\rm SI} &= \frac{4}{\pi A^2}\mu^2_r\frac{Y^2_1 \sin^2 2\theta_1}{M^4_h}\Big[\frac{m_p}{v}\Big(f^{p}_{Tu} + f^{p}_{Td} + f^{p}_{Ts} + \frac{2}{9}f^{p}_{TG}\\
			&+\frac{m_n}{v}\Big(f^{n}_{Tu} + f^{n}_{Td} + f^{n}_{Ts} + \frac{2}{9}f^{n}_{TG}\Big)\Big]^2
		\end{aligned}
	\end{equation}
	where A is the mass number of Xenon nucleus, $m_p (m_n)$ is mass of proton (neutron) and $M_h$ is mass of the SM Higgs boson. The coupling strengths of DM with quarks and gluons can be found in \cite{Dutta:2020xwn} and references therein.
 
The presence of the factor $Y^2_1 \sin^2 2\theta_1$ in the direct search cross-section given by Eq. \eqref{dda2}, explains why large mixing angle and large mass-splittings are ruled out from direct search constraints as $Y_1 = \Delta M_1 ~\sin2\theta_1/2 v$ makes the overall dependency of the direct search cross-section on the mass splitting and the singlet-doublet mixing as, $\sigma_{\rm SI} \propto \Delta M^2_1 \sin^42\theta_1 $. But as relic density requires larger $\sin\theta_1$ with large $\Delta M_1$ in order to be within the correct ballpark by virtue of large annihilations, thus the region roughly above $\Delta M_1 =$20 GeV can not simultaneously satisfy both the constraints. However, it is worth mentioning that at SM Higgs resonance ($M_{\chi_{1 3}}\sim M_h/2$) we can satisfy both relic density and direct search bound, where $\Delta M$ can be very large but with smaller $\sin\theta_1$ in the range $0.01 - 0.1$.

\begin{figure}[h!]
\centering
\includegraphics[scale=0.5]{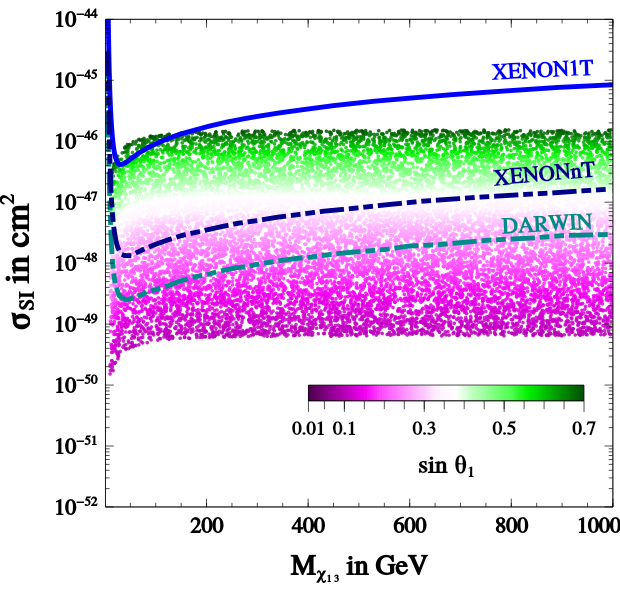}
\caption{Loop-induced spin-independent DM-nucleon cross-section function of DM mass.}
\label{lidd}
\end{figure} 

Even though tree level DM-nucleon scattering mediated by Z boson is forbidden due to Majorana nature of DM, one can have radiative spin-independent DM-nucleon scattering mediated by electroweak gauge bosons which, at one-loop level, gives the leading order contribution. One-loop SI DM-nucleon scattering cross-section can be estimated as~\cite{Bell:2018zra}
\begin{equation}
\label{ddloop}
\begin{aligned}
\sigma^{\rm loop}_{\rm SI} &= \frac{1}{\pi A^2}\mu^2_r \vert \mathcal{M} \vert^2
\end{aligned}
\end{equation}	
with amplitude,
\begin{equation}
\mathcal{M}= \frac{4 g^4 M_N M_{\chi_{1 3}}}{16 \pi^2 M^4_V} F\left(\frac{M^2_{\chi_{1 3}}}{M^2_V}\right) \sin^2\theta_1 \left[Z f_p + (A-Z) f_n\right]
\end{equation}
and the loop function $F$ is given by:
\begin{equation}
\label{amp}
\begin{aligned}
F(x)=&\frac{(8 x^2-4x+2)\log[\frac{\sqrt{1-4x}+1}{2\sqrt{x}}]}{4 x^2 \sqrt{1-4x}}\\+&\frac{\sqrt{1-4x}(2x+\log(x))}{4 x^2 \sqrt{1-4x}}
\end{aligned}
\end{equation}	
In the above expression, $M_{V}$ is the mass of SM vector boson ($W^\pm$ or $Z$), $M_N$ is nucleon mass and $f_p$ and $f_n$ are the interaction strengths (including hadronic uncertainties) of DM with proton and neutron respectively.  We assume conservation of isospin, {\it i.e.} $f_p/f_n = 1$. The value of $f_n$ varies within a range of $0.14<f_n<0.66$ and we take the central value $f_n \simeq 1/3$ ~\cite{Mei:2018qnt,Bhattacharya:2017sml,Cirelli:2005uq}. In Fig.~\ref{lidd}, we have shown this loop induced spin-independent DM-nucleon scattering cross-section as a function of DM mass $M_{\chi_{1 3}}$ where the colour map depicts the value of singlet-doublet mixing $\sin\theta_1$. Though this loop level interaction cross-section is almost safe from the existing constraints from XENON1T~\cite{Aprile:2018dbl}, future experiments like XENONnT~\cite{XENON:2015gkh} and DARWIN~\cite{DARWIN:2016hyl}, can probe the singlet-doublet mixing angle down to $0.1$, keeping the scenario verifiable.

\begin{figure}[t]
\includegraphics[scale=0.55]{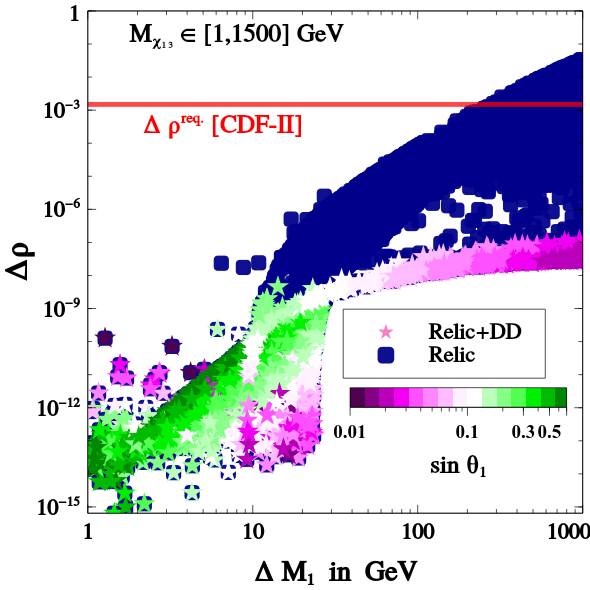}
\caption{$\Delta \rho$ is shown as a function of singlet-doublet mass splitting for the DM ($\Delta M_1$), along with the parameter space consistent with correct DM relic (Blue points) and relic + direct search constraints (coloured points) for the DM with a single generation of singlet-doublet fermions in the model.}\label{1gdelrho}
\end{figure}

\vspace{0.2cm}
\noindent
\textbf{\emph{W Boson Mass Anomaly}:} The $W$-boson mass has been precisely calculated within the SM framework in terms of input parameters $\{\alpha, G_F, M_Z\}$ which have been very accurately measured, with their numerical values given by \cite{ParticleDataGroup:2020ssz} 
\begin{eqnarray}\label{precise}
\alpha^{-1} &=& 137.035999084(51) \,, \,\,\delta \alpha / \alpha = 0.4 \times 10^{-9};\nonumber \\
 G_F &=& 1.1663787 \times 10^{-5} \, {\rm GeV}^{-2}, \,\, \delta G_F/G_F = 0.4 \times 10^{-5}; \nonumber\\
 M_{Z}&=& 91.1876 \pm 0.0021 \, {\rm GeV}, \,\, \delta M_{Z}/M_{Z} = 2.5\times 10^{-5}.\nonumber\\
\end{eqnarray}
The $W$-boson mass is related to these input parameters as~\cite{Hollik:1988ii, Nagashima:2010jma}
\begin{equation}
    M_W^2 \left(1-\frac{M_W^2}{ M_Z^2}\right) = \frac{\pi \alpha}{\sqrt{2} G_F}(1+\Delta r)
\end{equation}
where $\Delta r$ represents the contributions from the quantum corrections. Here $M_W$ and $M_Z$ are the renormalised masses in the on-shell scheme. From here $M_W$ can be calculated as
\begin{equation}
    M_W^2 = \frac{M_Z^2}{2}\left[ 1 + \sqrt{1- \frac{4 \pi \alpha}{ \sqrt{2} G_F M_Z^2}(1+\Delta r)} \right]~.
    \label{MW}
\end{equation}

\begin{figure*}[t]
	\centering
	\begin{tabular}{cc}
	\includegraphics[scale=0.5]{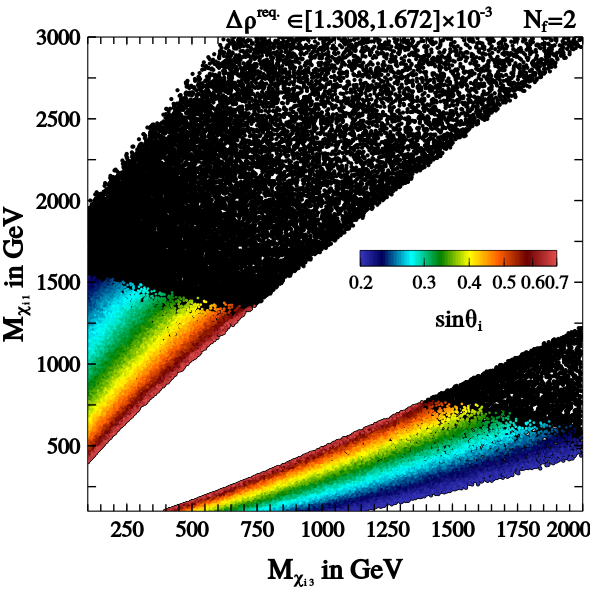}~~~~~~~~
	\includegraphics[scale=0.5]{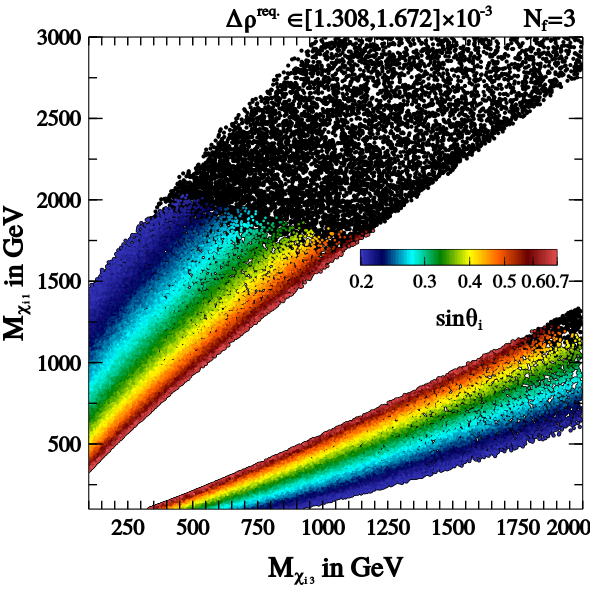}
	\end{tabular}
	\caption{Parameter space in the plane $M_{\chi_{i3}}$ and $M_{\chi_{i1}}$ giving $\Delta \rho^{\rm req}$ that can explain the CDF-II anomaly considering additional heavier fermion generations $N_f=2$ [Left panel] and $N_f=3$ [Right panel], while being consistent with the DM relevant constraints. The colour map shows the mixing between the singlet and doublet fermions of the heavier generations. The black points are obtained without imposing the constraint of perturbativity on the Yukawa coupling $Y_i$ {\it i.e.} $Y_{i} < \sqrt{4 \pi}$, where as once this constraint is considered the parameter space gets reduced as shown by the colored points.}
	\label{2g3gdelrho}
\end{figure*}

The radiative contribution $\Delta r$ can be written as \cite{Hollik:1988ii}
\begin{equation}
    \Delta r = \Delta \alpha -\frac{c_W^2}{s_W^2} \Delta \rho + \Delta r_{1}
\end{equation}
where $c_W=\cos{\theta_W}, s_W=\sin{\theta_W}$ with $\theta_W$ being the Weinberg angle. The main contribution to the $\Delta r$ is two-fold: one from pure QED
correction {\it i.e.} the change of the fine structure constant $\Delta \alpha$ evolved from $q^2= 0$ to $q^2 =M^2_Z$ and the other is $\Delta \rho$ which is the vacuum polarisation effect of the gauge boson generated by the top-bottom fermion loop.

The change in the fine structure constant $\Delta \alpha$ is
\begin{eqnarray}
\Delta \alpha &=& \frac{\alpha(M^2_Z)-\alpha(0)}{\alpha(0)}\nonumber\\
&=& -\frac{\alpha(0)}{3\pi} \sum_{m_f<M_Z} Q^2_f \Bigg[\frac{5}{3}-\ln(\frac{M^2_Z}{m^2_f})\Bigg]\nonumber\\
&=& 0.05943 \pm 0.00011
\end{eqnarray}
This arises from the renormalization of $\alpha$ which is dominated by light fermions.

Similarly $\Delta \rho$ is the oblique correction coming from non-negligible contribution due to the top and Higgs loop and is given by $(c_W^2/s_W^2)\Delta \rho \simeq 0.03$. In addition to these, there are contributions to $\Delta r$
from the vertex correction and the box diagrams which can be collectively written as $\Delta r_1= \Delta_{\rm Box+Vertex} \simeq 0.0064$.
If we use the central values of the parameters written in Eq.~\eqref{precise}, then we obtain that $\Delta r^{\rm}|_{\rm SM} = 0.0381387$, and consequently from Eq.\eqref{MW}, $M_W= 80.3564$ GeV. As this value of $M_W$ is $7\sigma$ below the value reported by CDF collaboration recently \cite{CDF:2022hxs}, the discrepancy may be resolved via quantum corrections that modify $\Delta r$ from its SM value. We find that a value of $\Delta r = 0.033192$ can lead to the central value from the CDF measurement {\it i.e.} $M_W = 80.4335$ GeV. As $\Delta r$ can get modified from oblique corrections, a new positive contribution to $\Delta \rho$ which we call $\Delta \rho^{\rm req} = 0.00149014$ is required to explain the W-mass anomaly. This positive contribution can come from the self-energy correction of the W-boson with the new fermion doublets in our setup. As the singlet and doublet fermions of each generation mix and give rise to the physical states $\chi_{ik}$, where index $i$ denotes the generation and $k$ denotes the particle mass eigenstate, the $\Delta \rho$ in our model can be calculated as as~\cite{Cynolter:2008ea}
\begin{widetext}
\begin{equation}
\Delta\rho=\sum_{i}\frac{1}{4 \pi^2 v^2}\left[2 \sin^2 \theta_i \cos^2 \theta_i ~\Pi(M_{\chi_{i3}},M_{\chi_{i1}},0)-2\cos^2 \theta_i ~\Pi(M_{i},M_{\chi_{i1}},0)-2\sin^2\theta_i ~\Pi(M_{i},M_{\chi_{i3}},0)\right]\,,
\end{equation}
 Here $\Pi(a,b,0)$ is given by
\begin{eqnarray}
\Pi (a,b,0) &=& -\frac{1}{2}(M_a^2+M_b^2)\left({\rm Div}+\ln \left(\frac{\mu_{ew}^2}{M_a M_b} \right)  \right)-\frac{1}{4} (M_a^2+ M_b^2)-\frac{(M_a^4+M_b^4)}
{4(M_a^2-M_b^2)} \ln \frac{M_b^2}{M_a^2} \nonumber\\
&& + M_a M_b \left\{{\rm Div}+ \ln  \left(\frac{\mu_{ew}^2}{M_a M_b} \right)+1+ \frac{(M_a^2 +M_b^2)}{2(M_a^2-M_b^2)} \ln  \frac{M_b^2}{M_a^2}  \right\} \,,
\end{eqnarray}
\end{widetext}
where Div$=\frac{1}{\epsilon} + \rm ln 4 \pi - \gamma_\epsilon $ that contains the divergent term in the dimensional regularisation method. Here it is worth mentioning that the values of the vacuum polarisations for identical masses ($M_b = M_a$) vanishes {\it i.e.} $\Pi(M_a,M_a,0)=0$ and hence $\chi_{i2}$ does not contribute to this oblique corrections as $M_{\chi_{i2}}=M_{\psi^{-}}=M_i$ which is clear from Eq.~\eqref{darkmass}. The $1\sigma$
allowed range for $\Delta \rho^{\rm req}$ consistent with the CDF $W$- Mass measurement is given by $\Delta \rho^{\rm req} : \{0.00130796,0.00167246\}$. Thus the explanation of W-mass anomaly constrains the model parameters space.

In Fig.~\ref{1gdelrho}, we show the value of $\Delta \rho$, as function of singlet-doublet mass splitting ($\Delta M_1$) such that the parameter space is consistent with correct relic and direct search constraints for the DM. The points satisfying correct relic density are shown by the blue colored square shaped points and the points satisfying both relic and direct detection constraints are shown by the other coloured star shaped points. The colour map represents the value of the mixing angle $\sin\theta_1$. The solid red line depicts the required value of the additional contribution to explain the CDF-II results. We find that to obtain the $\Delta \rho^{\rm req}$, we need a large mass splitting $\Delta M_1$ with $\sin\theta_1$ larger than $0.2$ or so. The fact that $\Delta \rho$ is sensitive to singlet doublet mass splitting $\Delta M_1$ is clear if we look at Eq.~\eqref{darkmass} and Eq.~\eqref{y1}, as the mass splitting between $\chi_{1 1}$ and $\psi^{-}_{1}$ namely, $\delta M_1$ can be expressed as $\delta M_1 = M_{\chi_1}-M_{\psi^{-}_{1}} = \Delta M_1 (1 - \cos^2{\theta_1})$. But as $\Delta M_{1}$ is severely constrained from direct search constraint except near the SM Higgs resonance, as shown in the right panel of Fig.~\ref{DMfig}, hence it is difficult to explain the W-mass anomaly while being consistent with the DM constraints. Even though for $M_{\chi{_{1 3}}} \sim M_{h}/2$, we see that large $\Delta M_1$ is still consistent from the relic and direct search point of view, but the fact that direct search bound constrains $\sin\theta$ upto $0.1$, keeps $\delta M_1$ small which is not sufficient to explain the W-Mass anomaly. Thus, single generation of singlet-doublet fermion, is not compatible from the perspective of W-mass anomaly, while being consistent with the DM relic and direct detection constraints. Incorporating additional singlet-doublet fermion generations will allow us to explain DM of the universe consistent with relic and direct detection constraints as well as explain the W-mass anomaly as the latter can be taken care of by the heavier generations. We have also found the impact of new physics on weak mixing angle $\theta_W$, by calculating the contribution of SD fermions to electroweak oblique parameters S, T, U \cite{Peskin:1991sw, Peskin:1990zt}, and using them to evaluate $\theta_W$ \cite{Kumar:2013yoa}. We find that our scenario can remain consistent with the measured value of $\theta_W$, keeping it within LEP ballpark. This is in contrast with models where W-boson anomaly is explained at tree level which changes only one of the oblique parameters, leading to significant deviation in $\theta_W$ from LEP measurement.

In Fig.~\ref{2g3gdelrho}, we have shown the parameter space consistent with $\Delta \rho^{\rm req}$ to explain the CDF-II anomaly in the plane $M_{\chi_{i3}}$ and $M_{\chi_{i1}}$ considering additional heavier fermion generations ($N_f=2$ for left panel and $N_f=3$ for the right panel). Here, we consider only the heavier 1 (left panel) and 2 (right panel) generations to be contributing to W-mass corrections while the lightest generation gives rise to DM phenomenology. 
Also, we ignore inter-generational mixing between singlet-doublet fermions for simplicity in this analysis such that $\theta$ corresponds to mixing within one generation only. For $N_f=3$ case, we consider identical mixing for the heavier two generations. Here it is worth mentioning that, as the Yukawa coupling $Y_{i}$ is proportional to $\Delta M_{i}$ and $\sin \theta_i$, some of the parameter space gets ruled out from the requirement of keeping this Yukawa coupling within the perturbative limit. We show the parameter space explaining the W mass anomaly without imposing the constraint of perturbativity by the black points where as the colored points are consistent with the constraint $Y_{i} < \sqrt{4 \pi}$. From Fig. \ref{2g3gdelrho}, it is also clear that if we consider additional fermion generations then we can obtain the $\Delta \rho^{\rm req}$ with smaller $\Delta M_{i}$. In addition to this, having $N_f=2$ or $N_f=3$ will also have consequence for neutrino mass as the former will lead to a vanishing lightest neutrino mass. While the $Z_2$-odd singlet scalar must be present from neutrino mass criteria, it does not play any role in W-mass correction. We also keep it sufficiently heavy so that its role in SD DM phenomenology via possible co-annihilations remain sub-dominant.

\vspace{0.2cm}
\noindent
\textbf{\emph{Conclusion}:} We have proposed a singlet-doublet fermion dark matter solution to the recently reported W boson mass anomaly by the CDF collaboration. The additional fermions can give rise to radiative correction to W boson mass to bring it closer to the CDF-II value, provided these new fermions have required mass as well as mass splittings. While single generation of singlet-doublet fermion can not satisfy both DM and W-mass requirements, including another generation of singlet-doublet fermion can lead to a successful scenario consistent with DM phenomenology and W-mass anomaly as the required W-mass correction can be generated dominantly by the heavier generation. We also find that our model remains consistent with the precision measurement of $\theta_W$ at LEP experiment while being able to explain W-boson mass anomaly. Multiple copies of singlet-doublet fermions can also lead to radiative origin of light neutrino masses if the model also contains a singlet scalar which, similar to singlet-doublet fermions, remains odd under an unbroken $Z_2$ symmetry. While independent confirmation of CDF anomaly can validate such scenarios further, the model also has several complementary detection prospects from colliders to dark matter direct detection keeping it verifiable in near future.

\vspace{0.2cm}

\acknowledgments
 NS would like to acknowledge the support from the Department of Atomic Energy (DAE)- Board of Research in Nuclear Sciences (BRNS), Government of India (Ref. Number: 58/14/15/2021- BRNS/37220).


\begin{thebibliography}{91}
	\expandafter\ifx\csname natexlab\endcsname\relax\def\natexlab#1{#1}\fi
	\expandafter\ifx\csname bibnamefont\endcsname\relax
	\def\bibnamefont#1{#1}\fi
	\expandafter\ifx\csname bibfnamefont\endcsname\relax
	\def\bibfnamefont#1{#1}\fi
	\expandafter\ifx\csname citenamefont\endcsname\relax
	\def\citenamefont#1{#1}\fi
	\expandafter\ifx\csname url\endcsname\relax
	\def\url#1{\texttt{#1}}\fi
	\expandafter\ifx\csname urlprefix\endcsname\relax\def\urlprefix{URL }\fi
	\providecommand{\bibinfo}[2]{#2}
	\providecommand{\eprint}[2][]{\url{#2}}
	
	\bibitem[{\citenamefont{Aaltonen et~al.}(2022)}]{CDF:2022hxs}
	\bibinfo{author}{\bibfnamefont{T.}~\bibnamefont{Aaltonen}} \bibnamefont{et~al.}
	(\bibinfo{collaboration}{CDF}), \bibinfo{journal}{Science}
	\textbf{\bibinfo{volume}{376}}, \bibinfo{pages}{170} (\bibinfo{year}{2022}).
	
	\bibitem[{\citenamefont{Fan et~al.}(2022{\natexlab{a}})\citenamefont{Fan, Li,
			Liu, and Lyu}}]{Fan:2022yly}
	\bibinfo{author}{\bibfnamefont{J.}~\bibnamefont{Fan}},
	\bibinfo{author}{\bibfnamefont{L.}~\bibnamefont{Li}},
	\bibinfo{author}{\bibfnamefont{T.}~\bibnamefont{Liu}}, \bibnamefont{and}
	\bibinfo{author}{\bibfnamefont{K.-F.} \bibnamefont{Lyu}}
	(\bibinfo{year}{2022}{\natexlab{a}}), \eprint{2204.04805}.
	
	\bibitem[{\citenamefont{Bagnaschi et~al.}(2022)\citenamefont{Bagnaschi, Ellis,
			Madigan, Mimasu, Sanz, and You}}]{Bagnaschi:2022whn}
	\bibinfo{author}{\bibfnamefont{E.}~\bibnamefont{Bagnaschi}},
	\bibinfo{author}{\bibfnamefont{J.}~\bibnamefont{Ellis}},
	\bibinfo{author}{\bibfnamefont{M.}~\bibnamefont{Madigan}},
	\bibinfo{author}{\bibfnamefont{K.}~\bibnamefont{Mimasu}},
	\bibinfo{author}{\bibfnamefont{V.}~\bibnamefont{Sanz}}, \bibnamefont{and}
	\bibinfo{author}{\bibfnamefont{T.}~\bibnamefont{You}} (\bibinfo{year}{2022}),
	\eprint{2204.05260}.
	
	\bibitem[{\citenamefont{de~Blas et~al.}(2022)\citenamefont{de~Blas, Pierini,
			Reina, and Silvestrini}}]{deBlas:2022hdk}
	\bibinfo{author}{\bibfnamefont{J.}~\bibnamefont{de~Blas}},
	\bibinfo{author}{\bibfnamefont{M.}~\bibnamefont{Pierini}},
	\bibinfo{author}{\bibfnamefont{L.}~\bibnamefont{Reina}}, \bibnamefont{and}
	\bibinfo{author}{\bibfnamefont{L.}~\bibnamefont{Silvestrini}}
	(\bibinfo{year}{2022}), \eprint{2204.04204}.
	
	\bibitem[{\citenamefont{Strumia}(2022)}]{Strumia:2022qkt}
	\bibinfo{author}{\bibfnamefont{A.}~\bibnamefont{Strumia}}
	(\bibinfo{year}{2022}), \eprint{2204.04191}.
	
	\bibitem[{\citenamefont{Asadi et~al.}(2022)\citenamefont{Asadi, Cesarotti,
			Fraser, Homiller, and Parikh}}]{Asadi:2022xiy}
	\bibinfo{author}{\bibfnamefont{P.}~\bibnamefont{Asadi}},
	\bibinfo{author}{\bibfnamefont{C.}~\bibnamefont{Cesarotti}},
	\bibinfo{author}{\bibfnamefont{K.}~\bibnamefont{Fraser}},
	\bibinfo{author}{\bibfnamefont{S.}~\bibnamefont{Homiller}}, \bibnamefont{and}
	\bibinfo{author}{\bibfnamefont{A.}~\bibnamefont{Parikh}}
	(\bibinfo{year}{2022}), \eprint{2204.05283}.
	
	\bibitem[{\citenamefont{Lu et~al.}(2022)\citenamefont{Lu, Wu, Wu, and
			Zhu}}]{Lu:2022bgw}
	\bibinfo{author}{\bibfnamefont{C.-T.} \bibnamefont{Lu}},
	\bibinfo{author}{\bibfnamefont{L.}~\bibnamefont{Wu}},
	\bibinfo{author}{\bibfnamefont{Y.}~\bibnamefont{Wu}}, \bibnamefont{and}
	\bibinfo{author}{\bibfnamefont{B.}~\bibnamefont{Zhu}} (\bibinfo{year}{2022}),
	\eprint{2204.03796}.
	
	\bibitem[{\citenamefont{Carpenter et~al.}(2022)\citenamefont{Carpenter, Murphy,
			and Smylie}}]{Carpenter:2022oyg}
	\bibinfo{author}{\bibfnamefont{L.~M.} \bibnamefont{Carpenter}},
	\bibinfo{author}{\bibfnamefont{T.}~\bibnamefont{Murphy}}, \bibnamefont{and}
	\bibinfo{author}{\bibfnamefont{M.~J.} \bibnamefont{Smylie}}
	(\bibinfo{year}{2022}), \eprint{2204.08546}.
	
	\bibitem[{\citenamefont{Fan et~al.}(2022{\natexlab{b}})\citenamefont{Fan, Tang,
			Tsai, and Wu}}]{Fan:2022dck}
	\bibinfo{author}{\bibfnamefont{Y.-Z.} \bibnamefont{Fan}},
	\bibinfo{author}{\bibfnamefont{T.-P.} \bibnamefont{Tang}},
	\bibinfo{author}{\bibfnamefont{Y.-L.~S.} \bibnamefont{Tsai}},
	\bibnamefont{and} \bibinfo{author}{\bibfnamefont{L.}~\bibnamefont{Wu}}
	(\bibinfo{year}{2022}{\natexlab{b}}), \eprint{2204.03693}.
	
	\bibitem[{\citenamefont{Zhu et~al.}(2022{\natexlab{a}})\citenamefont{Zhu, Cui,
			Xia, Yu, Huang, Yuan, and Fan}}]{Zhu:2022tpr}
	\bibinfo{author}{\bibfnamefont{C.-R.} \bibnamefont{Zhu}},
	\bibinfo{author}{\bibfnamefont{M.-Y.} \bibnamefont{Cui}},
	\bibinfo{author}{\bibfnamefont{Z.-Q.} \bibnamefont{Xia}},
	\bibinfo{author}{\bibfnamefont{Z.-H.} \bibnamefont{Yu}},
	\bibinfo{author}{\bibfnamefont{X.}~\bibnamefont{Huang}},
	\bibinfo{author}{\bibfnamefont{Q.}~\bibnamefont{Yuan}}, \bibnamefont{and}
	\bibinfo{author}{\bibfnamefont{Y.~Z.} \bibnamefont{Fan}}
	(\bibinfo{year}{2022}{\natexlab{a}}), \eprint{2204.03767}.
	
	\bibitem[{\citenamefont{Zhu et~al.}(2022{\natexlab{b}})\citenamefont{Zhu, Li,
			Cheng, Li, and Liang}}]{Zhu:2022scj}
	\bibinfo{author}{\bibfnamefont{B.-Y.} \bibnamefont{Zhu}},
	\bibinfo{author}{\bibfnamefont{S.}~\bibnamefont{Li}},
	\bibinfo{author}{\bibfnamefont{J.-G.} \bibnamefont{Cheng}},
	\bibinfo{author}{\bibfnamefont{R.-L.} \bibnamefont{Li}}, \bibnamefont{and}
	\bibinfo{author}{\bibfnamefont{Y.-F.} \bibnamefont{Liang}}
	(\bibinfo{year}{2022}{\natexlab{b}}), \eprint{2204.04688}.
	
	\bibitem[{\citenamefont{Kawamura et~al.}(2022)\citenamefont{Kawamura, Okawa,
			and Omura}}]{Kawamura:2022uft}
	\bibinfo{author}{\bibfnamefont{J.}~\bibnamefont{Kawamura}},
	\bibinfo{author}{\bibfnamefont{S.}~\bibnamefont{Okawa}}, \bibnamefont{and}
	\bibinfo{author}{\bibfnamefont{Y.}~\bibnamefont{Omura}}
	(\bibinfo{year}{2022}), \eprint{2204.07022}.
	
	\bibitem[{\citenamefont{Nagao et~al.}(2022)\citenamefont{Nagao, Nomura, and
			Okada}}]{Nagao:2022oin}
	\bibinfo{author}{\bibfnamefont{K.~I.} \bibnamefont{Nagao}},
	\bibinfo{author}{\bibfnamefont{T.}~\bibnamefont{Nomura}}, \bibnamefont{and}
	\bibinfo{author}{\bibfnamefont{H.}~\bibnamefont{Okada}}
	(\bibinfo{year}{2022}), \eprint{2204.07411}.
	
	\bibitem[{\citenamefont{Zhang and Feng}(2022)}]{Zhang:2022nnh}
	\bibinfo{author}{\bibfnamefont{K.-Y.} \bibnamefont{Zhang}} \bibnamefont{and}
	\bibinfo{author}{\bibfnamefont{W.-Z.} \bibnamefont{Feng}}
	(\bibinfo{year}{2022}), \eprint{2204.08067}.
	
	\bibitem[{\citenamefont{Liu et~al.}(2022)\citenamefont{Liu, Guo, Zhu, and
			Li}}]{Liu:2022jdq}
	\bibinfo{author}{\bibfnamefont{X.}~\bibnamefont{Liu}},
	\bibinfo{author}{\bibfnamefont{S.-Y.} \bibnamefont{Guo}},
	\bibinfo{author}{\bibfnamefont{B.}~\bibnamefont{Zhu}}, \bibnamefont{and}
	\bibinfo{author}{\bibfnamefont{Y.}~\bibnamefont{Li}} (\bibinfo{year}{2022}),
	\eprint{2204.04834}.
	
	\bibitem[{\citenamefont{Sakurai et~al.}(2022)\citenamefont{Sakurai, Takahashi,
			and Yin}}]{Sakurai:2022hwh}
	\bibinfo{author}{\bibfnamefont{K.}~\bibnamefont{Sakurai}},
	\bibinfo{author}{\bibfnamefont{F.}~\bibnamefont{Takahashi}},
	\bibnamefont{and} \bibinfo{author}{\bibfnamefont{W.}~\bibnamefont{Yin}}
	(\bibinfo{year}{2022}), \eprint{2204.04770}.
	
	\bibitem[{\citenamefont{Cacciapaglia and Sannino}(2022)}]{Cacciapaglia:2022xih}
	\bibinfo{author}{\bibfnamefont{G.}~\bibnamefont{Cacciapaglia}}
	\bibnamefont{and} \bibinfo{author}{\bibfnamefont{F.}~\bibnamefont{Sannino}}
	(\bibinfo{year}{2022}), \eprint{2204.04514}.
	
	\bibitem[{\citenamefont{Song et~al.}(2022)\citenamefont{Song, Su, and
			Zhang}}]{Song:2022xts}
	\bibinfo{author}{\bibfnamefont{H.}~\bibnamefont{Song}},
	\bibinfo{author}{\bibfnamefont{W.}~\bibnamefont{Su}}, \bibnamefont{and}
	\bibinfo{author}{\bibfnamefont{M.}~\bibnamefont{Zhang}}
	(\bibinfo{year}{2022}), \eprint{2204.05085}.
	
	\bibitem[{\citenamefont{Bahl et~al.}(2022)\citenamefont{Bahl, Braathen, and
			Weiglein}}]{Bahl:2022xzi}
	\bibinfo{author}{\bibfnamefont{H.}~\bibnamefont{Bahl}},
	\bibinfo{author}{\bibfnamefont{J.}~\bibnamefont{Braathen}}, \bibnamefont{and}
	\bibinfo{author}{\bibfnamefont{G.}~\bibnamefont{Weiglein}}
	(\bibinfo{year}{2022}), \eprint{2204.05269}.
	
	\bibitem[{\citenamefont{Cheng et~al.}(2022)\citenamefont{Cheng, He, Huang, and
			Li}}]{Cheng:2022jyi}
	\bibinfo{author}{\bibfnamefont{Y.}~\bibnamefont{Cheng}},
	\bibinfo{author}{\bibfnamefont{X.-G.} \bibnamefont{He}},
	\bibinfo{author}{\bibfnamefont{Z.-L.} \bibnamefont{Huang}}, \bibnamefont{and}
	\bibinfo{author}{\bibfnamefont{M.-W.} \bibnamefont{Li}}
	(\bibinfo{year}{2022}), \eprint{2204.05031}.
	
	\bibitem[{\citenamefont{Babu et~al.}(2022)\citenamefont{Babu, Jana, and
			K.}}]{Babu:2022pdn}
	\bibinfo{author}{\bibfnamefont{K.~S.} \bibnamefont{Babu}},
	\bibinfo{author}{\bibfnamefont{S.}~\bibnamefont{Jana}}, \bibnamefont{and}
	\bibinfo{author}{\bibfnamefont{V.~P.} \bibnamefont{K.}}
	(\bibinfo{year}{2022}), \eprint{2204.05303}.
	
	\bibitem[{\citenamefont{Heo et~al.}(2022)\citenamefont{Heo, Jung, and
			Lee}}]{Heo:2022dey}
	\bibinfo{author}{\bibfnamefont{Y.}~\bibnamefont{Heo}},
	\bibinfo{author}{\bibfnamefont{D.-W.} \bibnamefont{Jung}}, \bibnamefont{and}
	\bibinfo{author}{\bibfnamefont{J.~S.} \bibnamefont{Lee}}
	(\bibinfo{year}{2022}), \eprint{2204.05728}.
	
	\bibitem[{\citenamefont{Ahn et~al.}(2022)\citenamefont{Ahn, Kang, and
			Ramos}}]{Ahn:2022xeq}
	\bibinfo{author}{\bibfnamefont{Y.~H.} \bibnamefont{Ahn}},
	\bibinfo{author}{\bibfnamefont{S.~K.} \bibnamefont{Kang}}, \bibnamefont{and}
	\bibinfo{author}{\bibfnamefont{R.}~\bibnamefont{Ramos}}
	(\bibinfo{year}{2022}), \eprint{2204.06485}.
	
	\bibitem[{\citenamefont{Zheng et~al.}(2022)\citenamefont{Zheng, Chen, and
			Zhang}}]{Zheng:2022irz}
	\bibinfo{author}{\bibfnamefont{M.-D.} \bibnamefont{Zheng}},
	\bibinfo{author}{\bibfnamefont{F.-Z.} \bibnamefont{Chen}}, \bibnamefont{and}
	\bibinfo{author}{\bibfnamefont{H.-H.} \bibnamefont{Zhang}}
	(\bibinfo{year}{2022}), \eprint{2204.06541}.
	
	\bibitem[{\citenamefont{Perez et~al.}(2022)\citenamefont{Perez, Patel, and
			Plascencia}}]{Perez:2022uil}
	\bibinfo{author}{\bibfnamefont{P.~F.} \bibnamefont{Perez}},
	\bibinfo{author}{\bibfnamefont{H.~H.} \bibnamefont{Patel}}, \bibnamefont{and}
	\bibinfo{author}{\bibfnamefont{A.~D.} \bibnamefont{Plascencia}}
	(\bibinfo{year}{2022}), \eprint{2204.07144}.
	
	\bibitem[{\citenamefont{Kanemura and Yagyu}(2022)}]{Kanemura:2022ahw}
	\bibinfo{author}{\bibfnamefont{S.}~\bibnamefont{Kanemura}} \bibnamefont{and}
	\bibinfo{author}{\bibfnamefont{K.}~\bibnamefont{Yagyu}}
	(\bibinfo{year}{2022}), \eprint{2204.07511}.
	
	\bibitem[{\citenamefont{Borah et~al.}(2022{\natexlab{a}})\citenamefont{Borah,
			Mahapatra, Nanda, and Sahu}}]{Borah:2022obi}
	\bibinfo{author}{\bibfnamefont{D.}~\bibnamefont{Borah}},
	\bibinfo{author}{\bibfnamefont{S.}~\bibnamefont{Mahapatra}},
	\bibinfo{author}{\bibfnamefont{D.}~\bibnamefont{Nanda}}, \bibnamefont{and}
	\bibinfo{author}{\bibfnamefont{N.}~\bibnamefont{Sahu}}
	(\bibinfo{year}{2022}{\natexlab{a}}), \eprint{2204.08266}.
	
	\bibitem[{\citenamefont{Popov and Srivastava}(2022)}]{Popov:2022ldh}
	\bibinfo{author}{\bibfnamefont{O.}~\bibnamefont{Popov}} \bibnamefont{and}
	\bibinfo{author}{\bibfnamefont{R.}~\bibnamefont{Srivastava}}
	(\bibinfo{year}{2022}), \eprint{2204.08568}.
	
	\bibitem[{\citenamefont{Arcadi and Djouadi}(2022)}]{Arcadi:2022dmt}
	\bibinfo{author}{\bibfnamefont{G.}~\bibnamefont{Arcadi}} \bibnamefont{and}
	\bibinfo{author}{\bibfnamefont{A.}~\bibnamefont{Djouadi}}
	(\bibinfo{year}{2022}), \eprint{2204.08406}.
	
	\bibitem[{\citenamefont{Ghorbani and Ghorbani}(2022)}]{Ghorbani:2022vtv}
	\bibinfo{author}{\bibfnamefont{K.}~\bibnamefont{Ghorbani}} \bibnamefont{and}
	\bibinfo{author}{\bibfnamefont{P.}~\bibnamefont{Ghorbani}}
	(\bibinfo{year}{2022}), \eprint{2204.09001}.
	
	\bibitem[{\citenamefont{Han et~al.}(2022)\citenamefont{Han, Wang, Wang, Yang,
			and Zhang}}]{Han:2022juu}
	\bibinfo{author}{\bibfnamefont{X.-F.} \bibnamefont{Han}},
	\bibinfo{author}{\bibfnamefont{F.}~\bibnamefont{Wang}},
	\bibinfo{author}{\bibfnamefont{L.}~\bibnamefont{Wang}},
	\bibinfo{author}{\bibfnamefont{J.~M.} \bibnamefont{Yang}}, \bibnamefont{and}
	\bibinfo{author}{\bibfnamefont{Y.}~\bibnamefont{Zhang}}
	(\bibinfo{year}{2022}), \eprint{2204.06505}.
	
	\bibitem[{\citenamefont{Du et~al.}(2022{\natexlab{a}})\citenamefont{Du, Li,
			Wang, and Zhang}}]{Du:2022pbp}
	\bibinfo{author}{\bibfnamefont{X.~K.} \bibnamefont{Du}},
	\bibinfo{author}{\bibfnamefont{Z.}~\bibnamefont{Li}},
	\bibinfo{author}{\bibfnamefont{F.}~\bibnamefont{Wang}}, \bibnamefont{and}
	\bibinfo{author}{\bibfnamefont{Y.~K.} \bibnamefont{Zhang}}
	(\bibinfo{year}{2022}{\natexlab{a}}), \eprint{2204.04286}.
	
	\bibitem[{\citenamefont{Tang et~al.}(2022)\citenamefont{Tang, Abdughani, Feng,
			Tsai, and Fan}}]{Tang:2022pxh}
	\bibinfo{author}{\bibfnamefont{T.-P.} \bibnamefont{Tang}},
	\bibinfo{author}{\bibfnamefont{M.}~\bibnamefont{Abdughani}},
	\bibinfo{author}{\bibfnamefont{L.}~\bibnamefont{Feng}},
	\bibinfo{author}{\bibfnamefont{Y.-L.~S.} \bibnamefont{Tsai}},
	\bibnamefont{and} \bibinfo{author}{\bibfnamefont{Y.-Z.} \bibnamefont{Fan}}
	(\bibinfo{year}{2022}), \eprint{2204.04356}.
	
	\bibitem[{\citenamefont{Yang and Zhang}(2022)}]{Yang:2022gvz}
	\bibinfo{author}{\bibfnamefont{J.~M.} \bibnamefont{Yang}} \bibnamefont{and}
	\bibinfo{author}{\bibfnamefont{Y.}~\bibnamefont{Zhang}}
	(\bibinfo{year}{2022}), \eprint{2204.04202}.
	
	\bibitem[{\citenamefont{Athron et~al.}(2022{\natexlab{a}})\citenamefont{Athron,
			Bach, Jacob, Kotlarski, St\"ockinger, and Voigt}}]{Athron:2022isz}
	\bibinfo{author}{\bibfnamefont{P.}~\bibnamefont{Athron}},
	\bibinfo{author}{\bibfnamefont{M.}~\bibnamefont{Bach}},
	\bibinfo{author}{\bibfnamefont{D.~H.~J.} \bibnamefont{Jacob}},
	\bibinfo{author}{\bibfnamefont{W.}~\bibnamefont{Kotlarski}},
	\bibinfo{author}{\bibfnamefont{D.}~\bibnamefont{St\"ockinger}},
	\bibnamefont{and} \bibinfo{author}{\bibfnamefont{A.}~\bibnamefont{Voigt}}
	(\bibinfo{year}{2022}{\natexlab{a}}), \eprint{2204.05285}.
	
	\bibitem[{\citenamefont{Ghoshal et~al.}(2022)\citenamefont{Ghoshal, Okada,
			Okada, Raut, Shafi, and Thapa}}]{Ghoshal:2022vzo}
	\bibinfo{author}{\bibfnamefont{A.}~\bibnamefont{Ghoshal}},
	\bibinfo{author}{\bibfnamefont{N.}~\bibnamefont{Okada}},
	\bibinfo{author}{\bibfnamefont{S.}~\bibnamefont{Okada}},
	\bibinfo{author}{\bibfnamefont{D.}~\bibnamefont{Raut}},
	\bibinfo{author}{\bibfnamefont{Q.}~\bibnamefont{Shafi}}, \bibnamefont{and}
	\bibinfo{author}{\bibfnamefont{A.}~\bibnamefont{Thapa}}
	(\bibinfo{year}{2022}), \eprint{2204.07138}.
	
	\bibitem[{\citenamefont{Athron et~al.}(2022{\natexlab{b}})\citenamefont{Athron,
			Fowlie, Lu, Wu, Wu, and Zhu}}]{Athron:2022qpo}
	\bibinfo{author}{\bibfnamefont{P.}~\bibnamefont{Athron}},
	\bibinfo{author}{\bibfnamefont{A.}~\bibnamefont{Fowlie}},
	\bibinfo{author}{\bibfnamefont{C.-T.} \bibnamefont{Lu}},
	\bibinfo{author}{\bibfnamefont{L.}~\bibnamefont{Wu}},
	\bibinfo{author}{\bibfnamefont{Y.}~\bibnamefont{Wu}}, \bibnamefont{and}
	\bibinfo{author}{\bibfnamefont{B.}~\bibnamefont{Zhu}}
	(\bibinfo{year}{2022}{\natexlab{b}}), \eprint{2204.03996}.
	
	\bibitem[{\citenamefont{Blennow et~al.}(2022)\citenamefont{Blennow, Coloma,
			Fern\'andez-Mart\'\i{}nez, and Gonz\'alez-L\'opez}}]{Blennow:2022yfm}
	\bibinfo{author}{\bibfnamefont{M.}~\bibnamefont{Blennow}},
	\bibinfo{author}{\bibfnamefont{P.}~\bibnamefont{Coloma}},
	\bibinfo{author}{\bibfnamefont{E.}~\bibnamefont{Fern\'andez-Mart\'\i{}nez}},
	\bibnamefont{and}
	\bibinfo{author}{\bibfnamefont{M.}~\bibnamefont{Gonz\'alez-L\'opez}}
	(\bibinfo{year}{2022}), \eprint{2204.04559}.
	
	\bibitem[{\citenamefont{Heckman}(2022)}]{Heckman:2022the}
	\bibinfo{author}{\bibfnamefont{J.~J.} \bibnamefont{Heckman}}
	(\bibinfo{year}{2022}), \eprint{2204.05302}.
	
	\bibitem[{\citenamefont{Lee and Yamashita}(2022)}]{Lee:2022nqz}
	\bibinfo{author}{\bibfnamefont{H.~M.} \bibnamefont{Lee}} \bibnamefont{and}
	\bibinfo{author}{\bibfnamefont{K.}~\bibnamefont{Yamashita}}
	(\bibinfo{year}{2022}), \eprint{2204.05024}.
	
	\bibitem[{\citenamefont{Di~Luzio et~al.}(2022)\citenamefont{Di~Luzio, Gr\"ober,
			and Paradisi}}]{DiLuzio:2022xns}
	\bibinfo{author}{\bibfnamefont{L.}~\bibnamefont{Di~Luzio}},
	\bibinfo{author}{\bibfnamefont{R.}~\bibnamefont{Gr\"ober}}, \bibnamefont{and}
	\bibinfo{author}{\bibfnamefont{P.}~\bibnamefont{Paradisi}}
	(\bibinfo{year}{2022}), \eprint{2204.05284}.
	
	\bibitem[{\citenamefont{Paul and Valli}(2022)}]{Paul:2022dds}
	\bibinfo{author}{\bibfnamefont{A.}~\bibnamefont{Paul}} \bibnamefont{and}
	\bibinfo{author}{\bibfnamefont{M.}~\bibnamefont{Valli}}
	(\bibinfo{year}{2022}), \eprint{2204.05267}.
	
	\bibitem[{\citenamefont{Biek\"otter et~al.}(2022)\citenamefont{Biek\"otter,
			Heinemeyer, and Weiglein}}]{Biekotter:2022abc}
	\bibinfo{author}{\bibfnamefont{T.}~\bibnamefont{Biek\"otter}},
	\bibinfo{author}{\bibfnamefont{S.}~\bibnamefont{Heinemeyer}},
	\bibnamefont{and} \bibinfo{author}{\bibfnamefont{G.}~\bibnamefont{Weiglein}}
	(\bibinfo{year}{2022}), \eprint{2204.05975}.
	
	\bibitem[{\citenamefont{Balkin et~al.}(2022)\citenamefont{Balkin, Madge, Menzo,
			Perez, Soreq, and Zupan}}]{Balkin:2022glu}
	\bibinfo{author}{\bibfnamefont{R.}~\bibnamefont{Balkin}},
	\bibinfo{author}{\bibfnamefont{E.}~\bibnamefont{Madge}},
	\bibinfo{author}{\bibfnamefont{T.}~\bibnamefont{Menzo}},
	\bibinfo{author}{\bibfnamefont{G.}~\bibnamefont{Perez}},
	\bibinfo{author}{\bibfnamefont{Y.}~\bibnamefont{Soreq}}, \bibnamefont{and}
	\bibinfo{author}{\bibfnamefont{J.}~\bibnamefont{Zupan}}
	(\bibinfo{year}{2022}), \eprint{2204.05992}.
	
	\bibitem[{\citenamefont{Cheung et~al.}(2022)\citenamefont{Cheung, Keung, and
			Tseng}}]{Cheung:2022zsb}
	\bibinfo{author}{\bibfnamefont{K.}~\bibnamefont{Cheung}},
	\bibinfo{author}{\bibfnamefont{W.-Y.} \bibnamefont{Keung}}, \bibnamefont{and}
	\bibinfo{author}{\bibfnamefont{P.-Y.} \bibnamefont{Tseng}}
	(\bibinfo{year}{2022}), \eprint{2204.05942}.
	
	\bibitem[{\citenamefont{Du et~al.}(2022{\natexlab{b}})\citenamefont{Du, Li,
			Wang, and Zhang}}]{Du:2022brr}
	\bibinfo{author}{\bibfnamefont{X.~K.} \bibnamefont{Du}},
	\bibinfo{author}{\bibfnamefont{Z.}~\bibnamefont{Li}},
	\bibinfo{author}{\bibfnamefont{F.}~\bibnamefont{Wang}}, \bibnamefont{and}
	\bibinfo{author}{\bibfnamefont{Y.~K.} \bibnamefont{Zhang}}
	(\bibinfo{year}{2022}{\natexlab{b}}), \eprint{2204.05760}.
	
	\bibitem[{\citenamefont{Endo and Mishima}(2022)}]{Endo:2022kiw}
	\bibinfo{author}{\bibfnamefont{M.}~\bibnamefont{Endo}} \bibnamefont{and}
	\bibinfo{author}{\bibfnamefont{S.}~\bibnamefont{Mishima}}
	(\bibinfo{year}{2022}), \eprint{2204.05965}.
	
	\bibitem[{\citenamefont{Crivellin et~al.}(2022)\citenamefont{Crivellin, Kirk,
			Kitahara, and Mescia}}]{Crivellin:2022fdf}
	\bibinfo{author}{\bibfnamefont{A.}~\bibnamefont{Crivellin}},
	\bibinfo{author}{\bibfnamefont{M.}~\bibnamefont{Kirk}},
	\bibinfo{author}{\bibfnamefont{T.}~\bibnamefont{Kitahara}}, \bibnamefont{and}
	\bibinfo{author}{\bibfnamefont{F.}~\bibnamefont{Mescia}}
	(\bibinfo{year}{2022}), \eprint{2204.05962}.
	
	\bibitem[{\citenamefont{Mondal}(2022)}]{Mondal:2022xdy}
	\bibinfo{author}{\bibfnamefont{P.}~\bibnamefont{Mondal}}
	(\bibinfo{year}{2022}), \eprint{2204.07844}.
	
	\bibitem[{\citenamefont{Chowdhury et~al.}(2022)\citenamefont{Chowdhury, Heeck,
			Saad, and Thapa}}]{Chowdhury:2022moc}
	\bibinfo{author}{\bibfnamefont{T.~A.} \bibnamefont{Chowdhury}},
	\bibinfo{author}{\bibfnamefont{J.}~\bibnamefont{Heeck}},
	\bibinfo{author}{\bibfnamefont{S.}~\bibnamefont{Saad}}, \bibnamefont{and}
	\bibinfo{author}{\bibfnamefont{A.}~\bibnamefont{Thapa}}
	(\bibinfo{year}{2022}), \eprint{2204.08390}.
	
	\bibitem[{\citenamefont{Du et~al.}(2022{\natexlab{c}})\citenamefont{Du, Liu,
			and Nath}}]{Du:2022fqv}
	\bibinfo{author}{\bibfnamefont{M.}~\bibnamefont{Du}},
	\bibinfo{author}{\bibfnamefont{Z.}~\bibnamefont{Liu}}, \bibnamefont{and}
	\bibinfo{author}{\bibfnamefont{P.}~\bibnamefont{Nath}}
	(\bibinfo{year}{2022}{\natexlab{c}}), \eprint{2204.09024}.
	
	\bibitem[{\citenamefont{Bhaskar et~al.}(2022)\citenamefont{Bhaskar, Madathil,
			Mandal, and Mitra}}]{Bhaskar:2022vgk}
	\bibinfo{author}{\bibfnamefont{A.}~\bibnamefont{Bhaskar}},
	\bibinfo{author}{\bibfnamefont{A.~A.} \bibnamefont{Madathil}},
	\bibinfo{author}{\bibfnamefont{T.}~\bibnamefont{Mandal}}, \bibnamefont{and}
	\bibinfo{author}{\bibfnamefont{S.}~\bibnamefont{Mitra}}
	(\bibinfo{year}{2022}), \eprint{2204.09031}.
	
	\bibitem[{\citenamefont{Yuan et~al.}(2022)\citenamefont{Yuan, Zu, Feng, Cai,
			and Fan}}]{Yuan:2022cpw}
	\bibinfo{author}{\bibfnamefont{G.-W.} \bibnamefont{Yuan}},
	\bibinfo{author}{\bibfnamefont{L.}~\bibnamefont{Zu}},
	\bibinfo{author}{\bibfnamefont{L.}~\bibnamefont{Feng}},
	\bibinfo{author}{\bibfnamefont{Y.-F.} \bibnamefont{Cai}}, \bibnamefont{and}
	\bibinfo{author}{\bibfnamefont{Y.-Z.} \bibnamefont{Fan}}
	(\bibinfo{year}{2022}), \eprint{2204.04183}.
	
	\bibitem[{\citenamefont{Arias-Arag\'on
			et~al.}(2022)\citenamefont{Arias-Arag\'on, Fern\'andez-Mart\'\i{}nez,
			Gonz\'alez-L\'opez, and Merlo}}]{Arias-Aragon:2022ats}
	\bibinfo{author}{\bibfnamefont{F.}~\bibnamefont{Arias-Arag\'on}},
	\bibinfo{author}{\bibfnamefont{E.}~\bibnamefont{Fern\'andez-Mart\'\i{}nez}},
	\bibinfo{author}{\bibfnamefont{M.}~\bibnamefont{Gonz\'alez-L\'opez}},
	\bibnamefont{and} \bibinfo{author}{\bibfnamefont{L.}~\bibnamefont{Merlo}}
	(\bibinfo{year}{2022}), \eprint{2204.04672}.
	
	\bibitem[{\citenamefont{Mahbubani and Senatore}(2006)}]{Mahbubani:2005pt}
	\bibinfo{author}{\bibfnamefont{R.}~\bibnamefont{Mahbubani}} \bibnamefont{and}
	\bibinfo{author}{\bibfnamefont{L.}~\bibnamefont{Senatore}},
	\bibinfo{journal}{Phys. Rev.} \textbf{\bibinfo{volume}{D73}},
	\bibinfo{pages}{043510} (\bibinfo{year}{2006}), \eprint{hep-ph/0510064}.
	
	\bibitem[{\citenamefont{D'Eramo}(2007)}]{DEramo:2007anh}
	\bibinfo{author}{\bibfnamefont{F.}~\bibnamefont{D'Eramo}},
	\bibinfo{journal}{Phys. Rev.} \textbf{\bibinfo{volume}{D76}},
	\bibinfo{pages}{083522} (\bibinfo{year}{2007}), \eprint{0705.4493}.
	
	\bibitem[{\citenamefont{Enberg et~al.}(2007)\citenamefont{Enberg, Fox, Hall,
			Papaioannou, and Papucci}}]{Enberg:2007rp}
	\bibinfo{author}{\bibfnamefont{R.}~\bibnamefont{Enberg}},
	\bibinfo{author}{\bibfnamefont{P.~J.} \bibnamefont{Fox}},
	\bibinfo{author}{\bibfnamefont{L.~J.} \bibnamefont{Hall}},
	\bibinfo{author}{\bibfnamefont{A.~Y.} \bibnamefont{Papaioannou}},
	\bibnamefont{and} \bibinfo{author}{\bibfnamefont{M.}~\bibnamefont{Papucci}},
	\bibinfo{journal}{JHEP} \textbf{\bibinfo{volume}{11}}, \bibinfo{pages}{014}
	(\bibinfo{year}{2007}), \eprint{0706.0918}.
	
	\bibitem[{\citenamefont{Cohen et~al.}(2012)\citenamefont{Cohen, Kearney,
			Pierce, and Tucker-Smith}}]{Cohen:2011ec}
	\bibinfo{author}{\bibfnamefont{T.}~\bibnamefont{Cohen}},
	\bibinfo{author}{\bibfnamefont{J.}~\bibnamefont{Kearney}},
	\bibinfo{author}{\bibfnamefont{A.}~\bibnamefont{Pierce}}, \bibnamefont{and}
	\bibinfo{author}{\bibfnamefont{D.}~\bibnamefont{Tucker-Smith}},
	\bibinfo{journal}{Phys. Rev.} \textbf{\bibinfo{volume}{D85}},
	\bibinfo{pages}{075003} (\bibinfo{year}{2012}), \eprint{1109.2604}.
	
	\bibitem[{\citenamefont{Cheung and Sanford}(2014)}]{Cheung:2013dua}
	\bibinfo{author}{\bibfnamefont{C.}~\bibnamefont{Cheung}} \bibnamefont{and}
	\bibinfo{author}{\bibfnamefont{D.}~\bibnamefont{Sanford}},
	\bibinfo{journal}{JCAP} \textbf{\bibinfo{volume}{1402}}, \bibinfo{pages}{011}
	(\bibinfo{year}{2014}), \eprint{1311.5896}.
	
	\bibitem[{\citenamefont{Restrepo et~al.}(2015)\citenamefont{Restrepo, Rivera,
			Sánchez-Peláez, Zapata, and Tangarife}}]{Restrepo:2015ura}
	\bibinfo{author}{\bibfnamefont{D.}~\bibnamefont{Restrepo}},
	\bibinfo{author}{\bibfnamefont{A.}~\bibnamefont{Rivera}},
	\bibinfo{author}{\bibfnamefont{M.}~\bibnamefont{Sánchez-Peláez}},
	\bibinfo{author}{\bibfnamefont{O.}~\bibnamefont{Zapata}}, \bibnamefont{and}
	\bibinfo{author}{\bibfnamefont{W.}~\bibnamefont{Tangarife}},
	\bibinfo{journal}{Phys. Rev.} \textbf{\bibinfo{volume}{D92}},
	\bibinfo{pages}{013005} (\bibinfo{year}{2015}), \eprint{1504.07892}.
	
	\bibitem[{\citenamefont{Calibbi et~al.}(2015)\citenamefont{Calibbi, Mariotti,
			and Tziveloglou}}]{Calibbi:2015nha}
	\bibinfo{author}{\bibfnamefont{L.}~\bibnamefont{Calibbi}},
	\bibinfo{author}{\bibfnamefont{A.}~\bibnamefont{Mariotti}}, \bibnamefont{and}
	\bibinfo{author}{\bibfnamefont{P.}~\bibnamefont{Tziveloglou}},
	\bibinfo{journal}{JHEP} \textbf{\bibinfo{volume}{10}}, \bibinfo{pages}{116}
	(\bibinfo{year}{2015}), \eprint{1505.03867}.
	
	\bibitem[{\citenamefont{Cynolter et~al.}(2016)\citenamefont{Cynolter, Kovács,
			and Lendvai}}]{Cynolter:2015sua}
	\bibinfo{author}{\bibfnamefont{G.}~\bibnamefont{Cynolter}},
	\bibinfo{author}{\bibfnamefont{J.}~\bibnamefont{Kovács}}, \bibnamefont{and}
	\bibinfo{author}{\bibfnamefont{E.}~\bibnamefont{Lendvai}},
	\bibinfo{journal}{Mod. Phys. Lett.} \textbf{\bibinfo{volume}{A31}},
	\bibinfo{pages}{1650013} (\bibinfo{year}{2016}), \eprint{1509.05323}.
	
	\bibitem[{\citenamefont{Bhattacharya et~al.}(2016)\citenamefont{Bhattacharya,
			Sahoo, and Sahu}}]{Bhattacharya:2015qpa}
	\bibinfo{author}{\bibfnamefont{S.}~\bibnamefont{Bhattacharya}},
	\bibinfo{author}{\bibfnamefont{N.}~\bibnamefont{Sahoo}}, \bibnamefont{and}
	\bibinfo{author}{\bibfnamefont{N.}~\bibnamefont{Sahu}},
	\bibinfo{journal}{Phys. Rev. D} \textbf{\bibinfo{volume}{93}},
	\bibinfo{pages}{115040} (\bibinfo{year}{2016}), \eprint{1510.02760}.
	
	\bibitem[{\citenamefont{Bhattacharya
			et~al.}(2017{\natexlab{a}})\citenamefont{Bhattacharya, Sahoo, and
			Sahu}}]{Bhattacharya:2017sml}
	\bibinfo{author}{\bibfnamefont{S.}~\bibnamefont{Bhattacharya}},
	\bibinfo{author}{\bibfnamefont{N.}~\bibnamefont{Sahoo}}, \bibnamefont{and}
	\bibinfo{author}{\bibfnamefont{N.}~\bibnamefont{Sahu}},
	\bibinfo{journal}{Phys. Rev.} \textbf{\bibinfo{volume}{D96}},
	\bibinfo{pages}{035010} (\bibinfo{year}{2017}{\natexlab{a}}),
	\eprint{1704.03417}.
	
	\bibitem[{\citenamefont{Bhattacharya et~al.}(2018)\citenamefont{Bhattacharya,
			Ghosh, Sahoo, and Sahu}}]{Bhattacharya:2018fus}
	\bibinfo{author}{\bibfnamefont{S.}~\bibnamefont{Bhattacharya}},
	\bibinfo{author}{\bibfnamefont{P.}~\bibnamefont{Ghosh}},
	\bibinfo{author}{\bibfnamefont{N.}~\bibnamefont{Sahoo}}, \bibnamefont{and}
	\bibinfo{author}{\bibfnamefont{N.}~\bibnamefont{Sahu}}
	(\bibinfo{year}{2018}), \eprint{1812.06505}.
	
	\bibitem[{\citenamefont{Bhattacharya et~al.}(2019)\citenamefont{Bhattacharya,
			Ghosh, and Sahu}}]{Bhattacharya:2018cgx}
	\bibinfo{author}{\bibfnamefont{S.}~\bibnamefont{Bhattacharya}},
	\bibinfo{author}{\bibfnamefont{P.}~\bibnamefont{Ghosh}}, \bibnamefont{and}
	\bibinfo{author}{\bibfnamefont{N.}~\bibnamefont{Sahu}},
	\bibinfo{journal}{JHEP} \textbf{\bibinfo{volume}{02}}, \bibinfo{pages}{059}
	(\bibinfo{year}{2019}), \eprint{1809.07474}.
	
	\bibitem[{\citenamefont{Dutta~Banik et~al.}(2018)\citenamefont{Dutta~Banik,
			Saha, and Sil}}]{DuttaBanik:2018emv}
	\bibinfo{author}{\bibfnamefont{A.}~\bibnamefont{Dutta~Banik}},
	\bibinfo{author}{\bibfnamefont{A.~K.} \bibnamefont{Saha}}, \bibnamefont{and}
	\bibinfo{author}{\bibfnamefont{A.}~\bibnamefont{Sil}},
	\bibinfo{journal}{Phys. Rev.} \textbf{\bibinfo{volume}{D98}},
	\bibinfo{pages}{075013} (\bibinfo{year}{2018}), \eprint{1806.08080}.
	
	\bibitem[{\citenamefont{Barman et~al.}(2019{\natexlab{a}})\citenamefont{Barman,
			Bhattacharya, Ghosh, Kadam, and Sahu}}]{Barman:2019tuo}
	\bibinfo{author}{\bibfnamefont{B.}~\bibnamefont{Barman}},
	\bibinfo{author}{\bibfnamefont{S.}~\bibnamefont{Bhattacharya}},
	\bibinfo{author}{\bibfnamefont{P.}~\bibnamefont{Ghosh}},
	\bibinfo{author}{\bibfnamefont{S.}~\bibnamefont{Kadam}}, \bibnamefont{and}
	\bibinfo{author}{\bibfnamefont{N.}~\bibnamefont{Sahu}},
	\bibinfo{journal}{Phys. Rev. D} \textbf{\bibinfo{volume}{100}},
	\bibinfo{pages}{015027} (\bibinfo{year}{2019}{\natexlab{a}}),
	\eprint{1902.01217}.
	
	\bibitem[{\citenamefont{Bhattacharya
			et~al.}(2017{\natexlab{b}})\citenamefont{Bhattacharya, Karmakar, Sahu, and
			Sil}}]{Bhattacharya:2016rqj}
	\bibinfo{author}{\bibfnamefont{S.}~\bibnamefont{Bhattacharya}},
	\bibinfo{author}{\bibfnamefont{B.}~\bibnamefont{Karmakar}},
	\bibinfo{author}{\bibfnamefont{N.}~\bibnamefont{Sahu}}, \bibnamefont{and}
	\bibinfo{author}{\bibfnamefont{A.}~\bibnamefont{Sil}},
	\bibinfo{journal}{JHEP} \textbf{\bibinfo{volume}{05}}, \bibinfo{pages}{068}
	(\bibinfo{year}{2017}{\natexlab{b}}), \eprint{1611.07419}.
	
	\bibitem[{\citenamefont{Calibbi et~al.}(2018)\citenamefont{Calibbi,
			Lopez-Honorez, Lowette, and Mariotti}}]{Calibbi:2018fqf}
	\bibinfo{author}{\bibfnamefont{L.}~\bibnamefont{Calibbi}},
	\bibinfo{author}{\bibfnamefont{L.}~\bibnamefont{Lopez-Honorez}},
	\bibinfo{author}{\bibfnamefont{S.}~\bibnamefont{Lowette}}, \bibnamefont{and}
	\bibinfo{author}{\bibfnamefont{A.}~\bibnamefont{Mariotti}},
	\bibinfo{journal}{JHEP} \textbf{\bibinfo{volume}{09}}, \bibinfo{pages}{037}
	(\bibinfo{year}{2018}), \eprint{1805.04423}.
	
	\bibitem[{\citenamefont{Barman et~al.}(2019{\natexlab{b}})\citenamefont{Barman,
			Borah, Ghosh, and Saha}}]{Barman:2019aku}
	\bibinfo{author}{\bibfnamefont{B.}~\bibnamefont{Barman}},
	\bibinfo{author}{\bibfnamefont{D.}~\bibnamefont{Borah}},
	\bibinfo{author}{\bibfnamefont{P.}~\bibnamefont{Ghosh}}, \bibnamefont{and}
	\bibinfo{author}{\bibfnamefont{A.~K.} \bibnamefont{Saha}}
	(\bibinfo{year}{2019}{\natexlab{b}}), \eprint{1907.10071}.
	
	\bibitem[{\citenamefont{Dutta et~al.}(2021)\citenamefont{Dutta, Bhattacharya,
			Ghosh, and Sahu}}]{Dutta:2020xwn}
	\bibinfo{author}{\bibfnamefont{M.}~\bibnamefont{Dutta}},
	\bibinfo{author}{\bibfnamefont{S.}~\bibnamefont{Bhattacharya}},
	\bibinfo{author}{\bibfnamefont{P.}~\bibnamefont{Ghosh}}, \bibnamefont{and}
	\bibinfo{author}{\bibfnamefont{N.}~\bibnamefont{Sahu}},
	\bibinfo{journal}{JCAP} \textbf{\bibinfo{volume}{03}}, \bibinfo{pages}{008}
	(\bibinfo{year}{2021}), \eprint{2009.00885}.
	
	\bibitem[{\citenamefont{Borah et~al.}(2022{\natexlab{b}})\citenamefont{Borah,
			Dutta, Mahapatra, and Sahu}}]{Borah:2021khc}
	\bibinfo{author}{\bibfnamefont{D.}~\bibnamefont{Borah}},
	\bibinfo{author}{\bibfnamefont{M.}~\bibnamefont{Dutta}},
	\bibinfo{author}{\bibfnamefont{S.}~\bibnamefont{Mahapatra}},
	\bibnamefont{and} \bibinfo{author}{\bibfnamefont{N.}~\bibnamefont{Sahu}},
	\bibinfo{journal}{Phys. Rev. D} \textbf{\bibinfo{volume}{105}},
	\bibinfo{pages}{015029} (\bibinfo{year}{2022}{\natexlab{b}}),
	\eprint{2109.02699}.
	
	\bibitem[{\citenamefont{Borah et~al.}(2021)\citenamefont{Borah, Dutta,
			Mahapatra, and Sahu}}]{Borah:2021rbx}
	\bibinfo{author}{\bibfnamefont{D.}~\bibnamefont{Borah}},
	\bibinfo{author}{\bibfnamefont{M.}~\bibnamefont{Dutta}},
	\bibinfo{author}{\bibfnamefont{S.}~\bibnamefont{Mahapatra}},
	\bibnamefont{and} \bibinfo{author}{\bibfnamefont{N.}~\bibnamefont{Sahu}}
	(\bibinfo{year}{2021}), \eprint{2112.06847}.
	
	\bibitem[{\citenamefont{Fraser et~al.}(2014)\citenamefont{Fraser, Ma, and
			Popov}}]{Fraser:2014yha}
	\bibinfo{author}{\bibfnamefont{S.}~\bibnamefont{Fraser}},
	\bibinfo{author}{\bibfnamefont{E.}~\bibnamefont{Ma}}, \bibnamefont{and}
	\bibinfo{author}{\bibfnamefont{O.}~\bibnamefont{Popov}},
	\bibinfo{journal}{Phys. Lett. B} \textbf{\bibinfo{volume}{737}},
	\bibinfo{pages}{280} (\bibinfo{year}{2014}), \eprint{1408.4785}.
	
	\bibitem[{\citenamefont{Konar et~al.}(2020)\citenamefont{Konar, Mukherjee,
			Saha, and Show}}]{Konar:2020wvl}
	\bibinfo{author}{\bibfnamefont{P.}~\bibnamefont{Konar}},
	\bibinfo{author}{\bibfnamefont{A.}~\bibnamefont{Mukherjee}},
	\bibinfo{author}{\bibfnamefont{A.~K.} \bibnamefont{Saha}}, \bibnamefont{and}
	\bibinfo{author}{\bibfnamefont{S.}~\bibnamefont{Show}},
	\bibinfo{journal}{Phys. Rev. D} \textbf{\bibinfo{volume}{102}},
	\bibinfo{pages}{015024} (\bibinfo{year}{2020}), \eprint{2001.11325}.
	
	\bibitem[{\citenamefont{Belanger et~al.}(2009)\citenamefont{Belanger, Boudjema,
			Pukhov, and Semenov}}]{Belanger:2008sj}
	\bibinfo{author}{\bibfnamefont{G.}~\bibnamefont{Belanger}},
	\bibinfo{author}{\bibfnamefont{F.}~\bibnamefont{Boudjema}},
	\bibinfo{author}{\bibfnamefont{A.}~\bibnamefont{Pukhov}}, \bibnamefont{and}
	\bibinfo{author}{\bibfnamefont{A.}~\bibnamefont{Semenov}},
	\bibinfo{journal}{Comput. Phys. Commun.} \textbf{\bibinfo{volume}{180}},
	\bibinfo{pages}{747} (\bibinfo{year}{2009}), \eprint{0803.2360}.
	
	\bibitem[{\citenamefont{Aghanim et~al.}(2018)}]{Aghanim:2018eyx}
	\bibinfo{author}{\bibfnamefont{N.}~\bibnamefont{Aghanim}} \bibnamefont{et~al.}
	(\bibinfo{collaboration}{Planck}) (\bibinfo{year}{2018}),
	\eprint{1807.06209}.
	
	\bibitem[{\citenamefont{Aprile et~al.}(2018)}]{Aprile:2018dbl}
	\bibinfo{author}{\bibfnamefont{E.}~\bibnamefont{Aprile}} \bibnamefont{et~al.}
	(\bibinfo{year}{2018}), \eprint{1805.12562}.
	
	\bibitem[{\citenamefont{Bell et~al.}(2018)\citenamefont{Bell, Busoni, and
			Sanderson}}]{Bell:2018zra}
	\bibinfo{author}{\bibfnamefont{N.~F.} \bibnamefont{Bell}},
	\bibinfo{author}{\bibfnamefont{G.}~\bibnamefont{Busoni}}, \bibnamefont{and}
	\bibinfo{author}{\bibfnamefont{I.~W.} \bibnamefont{Sanderson}},
	\bibinfo{journal}{JCAP} \textbf{\bibinfo{volume}{08}}, \bibinfo{pages}{017}
	(\bibinfo{year}{2018}), \bibinfo{note}{[Erratum: JCAP 01, E01 (2019)]},
	\eprint{1803.01574}.
	
	\bibitem[{\citenamefont{Mei and Wei}(2018)}]{Mei:2018qnt}
	\bibinfo{author}{\bibfnamefont{D.~M.} \bibnamefont{Mei}} \bibnamefont{and}
	\bibinfo{author}{\bibfnamefont{W.~Z.} \bibnamefont{Wei}},
	\bibinfo{journal}{Phys. Lett. B} \textbf{\bibinfo{volume}{785}},
	\bibinfo{pages}{610} (\bibinfo{year}{2018}).
	
	\bibitem[{\citenamefont{Cirelli et~al.}(2006)\citenamefont{Cirelli, Fornengo,
			and Strumia}}]{Cirelli:2005uq}
	\bibinfo{author}{\bibfnamefont{M.}~\bibnamefont{Cirelli}},
	\bibinfo{author}{\bibfnamefont{N.}~\bibnamefont{Fornengo}}, \bibnamefont{and}
	\bibinfo{author}{\bibfnamefont{A.}~\bibnamefont{Strumia}},
	\bibinfo{journal}{Nucl. Phys.} \textbf{\bibinfo{volume}{B753}},
	\bibinfo{pages}{178} (\bibinfo{year}{2006}), \eprint{hep-ph/0512090}.
	
	\bibitem[{\citenamefont{Aprile et~al.}(2016)}]{XENON:2015gkh}
	\bibinfo{author}{\bibfnamefont{E.}~\bibnamefont{Aprile}} \bibnamefont{et~al.}
	(\bibinfo{collaboration}{XENON}), \bibinfo{journal}{JCAP}
	\textbf{\bibinfo{volume}{04}}, \bibinfo{pages}{027} (\bibinfo{year}{2016}),
	\eprint{1512.07501}.
	
	\bibitem[{\citenamefont{Aalbers et~al.}(2016)}]{DARWIN:2016hyl}
	\bibinfo{author}{\bibfnamefont{J.}~\bibnamefont{Aalbers}} \bibnamefont{et~al.}
	(\bibinfo{collaboration}{DARWIN}), \bibinfo{journal}{JCAP}
	\textbf{\bibinfo{volume}{11}}, \bibinfo{pages}{017} (\bibinfo{year}{2016}),
	\eprint{1606.07001}.
	
	\bibitem[{\citenamefont{Zyla et~al.}(2020)}]{ParticleDataGroup:2020ssz}
	\bibinfo{author}{\bibfnamefont{P.~A.} \bibnamefont{Zyla}} \bibnamefont{et~al.}
	(\bibinfo{collaboration}{Particle Data Group}), \bibinfo{journal}{PTEP}
	\textbf{\bibinfo{volume}{2020}}, \bibinfo{pages}{083C01}
	(\bibinfo{year}{2020}).
	
	\bibitem[{\citenamefont{Hollik}(1990)}]{Hollik:1988ii}
	\bibinfo{author}{\bibfnamefont{W.~F.~L.} \bibnamefont{Hollik}},
	\bibinfo{journal}{Fortsch. Phys.} \textbf{\bibinfo{volume}{38}},
	\bibinfo{pages}{165} (\bibinfo{year}{1990}).
	
	\bibitem[{\citenamefont{Nagashima}(2010)}]{Nagashima:2010jma}
	\bibinfo{author}{\bibfnamefont{Y.}~\bibnamefont{Nagashima}},
	\emph{\bibinfo{title}{{Elementary particle physics: Foundations of the
				standard model, volume 2}}} (\bibinfo{publisher}{Wiley-VCH},
	\bibinfo{address}{Weinheim}, \bibinfo{year}{2010}), ISBN
	\bibinfo{isbn}{978-3-527-40966-2}.
	
	\bibitem[{\citenamefont{Cynolter and Lendvai}(2008)}]{Cynolter:2008ea}
	\bibinfo{author}{\bibfnamefont{G.}~\bibnamefont{Cynolter}} \bibnamefont{and}
	\bibinfo{author}{\bibfnamefont{E.}~\bibnamefont{Lendvai}},
	\bibinfo{journal}{Eur. Phys. J.} \textbf{\bibinfo{volume}{C58}},
	\bibinfo{pages}{463} (\bibinfo{year}{2008}), \eprint{0804.4080}.
	
	\bibitem[{\citenamefont{Peskin and Takeuchi}(1992)}]{Peskin:1991sw}
	\bibinfo{author}{\bibfnamefont{M.~E.} \bibnamefont{Peskin}} \bibnamefont{and}
	\bibinfo{author}{\bibfnamefont{T.}~\bibnamefont{Takeuchi}},
	\bibinfo{journal}{Phys. Rev.} \textbf{\bibinfo{volume}{D46}},
	\bibinfo{pages}{381} (\bibinfo{year}{1992}).
	
	\bibitem[{\citenamefont{Peskin and Takeuchi}(1990)}]{Peskin:1990zt}
	\bibinfo{author}{\bibfnamefont{M.~E.} \bibnamefont{Peskin}} \bibnamefont{and}
	\bibinfo{author}{\bibfnamefont{T.}~\bibnamefont{Takeuchi}},
	\bibinfo{journal}{Phys. Rev. Lett.} \textbf{\bibinfo{volume}{65}},
	\bibinfo{pages}{964} (\bibinfo{year}{1990}).
	
	\bibitem[{\citenamefont{Kumar et~al.}(2013)\citenamefont{Kumar, Mantry,
			Marciano, and Souder}}]{Kumar:2013yoa}
	\bibinfo{author}{\bibfnamefont{K.~S.} \bibnamefont{Kumar}},
	\bibinfo{author}{\bibfnamefont{S.}~\bibnamefont{Mantry}},
	\bibinfo{author}{\bibfnamefont{W.~J.} \bibnamefont{Marciano}},
	\bibnamefont{and} \bibinfo{author}{\bibfnamefont{P.~A.}
		\bibnamefont{Souder}}, \bibinfo{journal}{Ann. Rev. Nucl. Part. Sci.}
	\textbf{\bibinfo{volume}{63}}, \bibinfo{pages}{237} (\bibinfo{year}{2013}),
	\eprint{1302.6263}.
	
\end{thebibliography}
%

\end{document}